\documentclass[12pt]{article}  
\usepackage{a41}
\usepackage{cite}
\usepackage{amsmath}
\usepackage{amssymb}
\usepackage{color}  

\usepackage[rflt]{floatflt}              
\usepackage{float}
\usepackage{slashed}
\def\b0{\beta_0}

\usepackage{array}

\usepackage[english]{babel}
\usepackage[latin1]{inputenc}
\usepackage[T1]{fontenc}
\usepackage{ae}

\usepackage{url}

\usepackage{amsmath, amsthm, amssymb}
\newtheorem{thm}{Theorem}[section]

\newtheorem{definition}[thm]{Definition}

\newcommand{\Li}{{\rm Li}}

\newcommand{\ep}{\varepsilon}

\usepackage{rotating}

\usepackage{graphicx}
\newcounter{mmacnt}
\def\restartmma{\setcounter{mmacnt}{0}}
\restartmma \catcode`|=\active
\def|#1|{\mathrm{#1}}
\catcode`|=12
\newenvironment{mma}{
 \par\smallskip
 \catcode`|=\active
 \parskip=0pt\parindent=0pt % locally
 \small
 \def\In##1\\{%
\def\linebreak{\hfill\break\null\qquad}%
\refstepcounter{mmacnt}
\hangindent=2.5em\hangafter=0
\leavevmode
\llap{\tiny\sffamily n[\arabic{mmacnt}]:=\kern.5em}%
\mathversion{bold}\footnotesize$\displaystyle##1$\normalsize
\mathversion{normal}\par
 }%
 \def\Print##1\\{%
\def\linebreak{\hfill\break}%
\hangindent=2.5em\hangafter=0
\leavevmode ##1\par}%
 \def\Out##1\\{%
\def\linebreak{$\hfill\break\null\hfill$}%
\kern\abovedisplayskip\par
\hangindent=2.5em\hangafter=0
\leavevmode
\llap{\tiny\sffamily Out[\arabic{mmacnt}]=\kern.5em}
\footnotesize$\displaystyle##1$\normalsize\hfill\null\par
\kern\belowdisplayskip
 }%
 \def\Warning##1##2\\{%
\def\linebreak{\hfill\break}%
\hangindent=2.5em\hangafter=0
\leavevmode
{\scriptsize##1 : ##2}\par}%
}{%
 \par\smallskip
}

\usepackage{color}

\newenvironment{fshaded}{%
\MakeFramed {\FrameRestore}
}%
{\endMakeFramed}

\def\b0{\beta_0}

\def\cC{{\cal C}}

\def\Gp0{{\Gamma^{'}_0}}
\def\Gp1{{\Gamma^{'}_1}}
\def\Gp2{{\Gamma^{'}_2}}

\allowdisplaybreaks[4]

\begin{document}
\setlength{\baselineskip}{0.515cm}

\sloppy
\thispagestyle{empty}
\begin{flushleft}
DESY 18--163
%\hfill {\tt arXiv:1810.xxxxx[hep-ph]}
\\
DO--TH 18/24\\
TIF-UNIMI-2018-7
\end{flushleft}

\mbox{}
\vspace*{\fill}
\begin{center}

{\LARGE\bf The asymptotic behavior of the heavy quark}

\vspace*{3mm} 
{\LARGE\bf form factors at higher order}

\vspace{3cm}
\large
{\large 
J.~Bl\"umlein$^a$, 
P.~Marquard$^a$,
and
N.~Rana$^{a,b}$
}

\vspace{1.cm}
\normalsize
{\it   $^a$~Deutsches Elektronen--Synchrotron, DESY,}\\
{\it   Platanenallee 6, D-15738 Zeuthen, Germany}

\vspace*{2mm}
{\it  $^b$~INFN, Sezione di Milano, Via Celoria 16, I-20133 Milano, Italy}

%%\today

\end{center}
\normalsize
\vspace{\fill}
\begin{abstract}
\noindent
In the asymptotic limit $Q^2 \gg m^2$, the heavy quark form factors exhibit Sudakov behavior. 
We study the corresponding renormalization group equations of the heavy quark form factors which 
do not only govern the structure of infrared divergences, but also control the high energy 
logarithms. This enables us to obtain the complete logarithmic three--loop and partial 
four--loop contributions to the heavy quark form factors in perturbative Quantum Chromodynamics.
\end{abstract}

\vspace*{\fill}
\noindent
% \numberwithin{equation}{section}
%%%%%%%%%%%%%%%%%%%%%%%%%%%%%%%%%%%%%%%%%%%%%%%%%%%%%%%%%%%%%%%%%%%%%%%%%%%%%%%%%%%%%%%%%%%%%%%%%%%%%%%%%%%%%%%%%%%%%%%%%%%%%%%%%%%
\newpage

%--------------------------------------------------------------------------------------------------------
\section{Introduction}
\label{sec:1}
%--------------------------------------------------------------------------------------------------------

\vspace*{1mm}
\noindent
Amplitudes for hard scattering processes in Quantum Chromodynamics (QCD) at higher order do provide 
precise phenomenological predictions for scattering processes and also a clear insight into underlying 
principles such as factorization or the universality of infrared (IR) singularities. The latter 
properties can be used to resum large logarithmic contributions by applying the corresponding evolution 
equations either globally or in particular kinematic regions. Especially for massless scattering 
amplitudes, remarkable progress has been made in the understanding of the structure of IR divergences 
due to the interplay of the soft- and collinear dynamics. A first step was taken in \cite{Catani:1998bh}, 
where a prediction for the singularities of two--loop amplitudes was given except the single pole in 
the dimensional variable  $\ep = (4-D)/2$. Later, generalizations of this result for multi-parton
amplitudes have been obtained in \cite{Sterman:2002qn, Becher:2009cu, Gardi:2009qi} beyond two--loop
order. The IR structure is more prominent and interesting especially in the case of the form factors.
The interplay of the soft and collinear anomalous dimensions building up the singular structure of 
the form factors has first been noticed in \cite{Ravindran:2004mb} at two--loop order and has been 
later established at three--loop order in \cite{Moch:2005tm}. In many following publications
this has been studied in detail, see e.g. \cite{Rana:2016vlw} for references.

It is also of interest to generalize these considerations to the massive  case. Here a first step has 
been taken in \cite{Catani:2000ef} by obtaining the IR singularities for one--loop scattering amplitudes 
with massive partons. Next, in the asymptotic limit $Q^2 \gg m^2$, a factorization theorem has been 
proposed in \cite{Penin:2005eh, Mitov:2006xs, Becher:2007cu}. While a first step was taken in 
\cite{Mitov:2009sv} to obtain the IR structure of a generic two--loop amplitude without considering the 
small mass limit, finally in \cite{Becher:2009kw} the general solution has been presented for the singular 
structure of a generic scattering amplitude containing massless and massive partons.

The universality of these IR singularities, along with the factorization of QCD amplitudes, presents a 
rich structure even for the massive case. Especially in the asymptotic limit, the amplitudes with massive 
partons exhibit the Sudakov behavior. It was first studied in \cite{Mitov:2006xs} and a general 
factorization formula was presented. Recently a partial result for the poles in $\ep$ of the heavy quark 
vector form factor in the leading color limit has been obtained up to four--loop order in \cite{Ahmed:2017gyt} 
solving the renormalization group equation (RGE). 

In the present paper, we study this behavior of the different heavy quark form factors following the method proposed for 
massless cases in \cite{Ravindran:2005vv, Ravindran:2006cg} and in the massive case of 
Ref.~\cite{Mitov:2006xs}. From a phenomenological 
perspective, the 
closer understanding of heavy quark production plays a significant role in current studies in elementary
particle physics, from precision measurements of the Standard Model parameters to the search for signals 
from beyond the Standard Model. Henceforth, there has been much attention during last decade to obtain 
precise theoretical predictions for physical quantities involving heavy quarks, see also \cite{Ablinger:2010ty,
Ablinger:2014vwa,Ablinger:2014nga,Ablinger:2014lka,AGG,Behring:2014eya,Ablinger:2017err,Ablinger:2017ptf}.
Important elements to all 
these predictions are the form factors. Dedicated work is going on for decades to obtain the heavy quark 
form factors for different currents, namely vector, axial-vector, scalar and pseudo-scalar currents at 
two--loop order \cite{Bernreuther:2004ih,Bernreuther:2004th,Bernreuther:2005rw,Bernreuther:2005gw,
Gluza:2009yy,Ablinger:2017hst} and three--loop order \cite{Henn:2016tyf, Ablinger:2018yae, FORMF3, 
Lee:2018rgs} in perturbative QCD. Since the complete computations are very challenging, results obtained 
in certain kinematic limits form important checks to these calculations. 

We present the complete three--loop results for the form factors in the asymptotic limit, retaining all 
logarithmic orders. For the ${\cal O}(\ep^0 L^0)$ terms we obtain the leading color contributions. A partial result 
for the vector form factor has been given in \cite{Gluza:2009yy} for three--loop order, where unknown 
coefficients $K_I$ at three-loop order were yet missing. We also present yet approximate four--loop results, 
combining all ingredients known at present and line out which missing terms have still to be calculated.
%--------------------------------------------------------------------------------------------------------
\section{The Sudakov behavior} 
\label{sec:thr}
%--------------------------------------------------------------------------------------------------------

\vspace*{1mm}
\noindent
We consider the renormalized form factors $F_{I}$ arising in the decay of a colorless massive boson, $I$, of 
momentum $Q$ to a pair of heavy quarks of mass $m$ in the asymptotic limit $Q^2 \gg m^2$. Here $Q^2$ is the 
center-of-mass energy squared and $I=V,A,S$ and $P$ indicates a vector, an axial-vector, a scalar and a 
pseudo-scalar boson, respectively. In the asymptotic limit the differences between the  vector and 
axial-vector form factors and between scalar and pseudo-scalar form factors vanish. Also the magnetic form 
factor in the vector and axial-vector case vanish in this limit. Henceforth, we therefore only consider the 
electric form factor ($F_V$) for the vector current and the scalar form factor ($F_S$), 
%--------------------------------------------------------------------------------------------------------
\begin{align}
& -i v_Q \delta^{cd} \gamma^\mu F_V^{cd} \\
& - \frac{m}{v} s_Q \delta_{cd} F_S^{cd},
\end{align}
%--------------------------------------------------------------------------------------------------------
where $v = (\sqrt{2} G_F)^{-1/2}$  is the vacuum expectation value, $v_Q$ and $s_Q$ are the heavy
quark vector and scalar couplings, and $c,d$ are color indices. We consider $n_l$ light 
quarks and a single heavy quark and deal with the case of only massless QCD contributions to the heavy quark 
form factors, i.e. no internal massive lines.

In the asymptotic limit, the functions  $\hat{F}_I \left(a_s(\mu), \frac{Q^2}{\mu^2}, \frac{m^2}{\mu^2}, \ep 
\right)$ 
satisfy the following 
integro-differential equation \cite{COLLINS} 
%--------------------------------------------------------------------------------------------------------
\begin{equation} \label{eq:kge}
 \mu^2 \frac{\partial}{\partial \mu^2} \ln \hat{F}_{I}\left(\frac{Q^2}{\mu^2},\frac{m^2}{\mu^2}, a_s,\ep\right) 
= \frac{1}{2} \left[ K_{I}\left(\frac{m^2}{\mu^2},a_s,\ep\right) + 
G_{I}\left(\frac{Q^2}{\mu^2},a_s,\ep\right) \right].
\end{equation}
%--------------------------------------------------------------------------------------------------------
Here $\hat{F}_I$ contains all logarithmic and infrared pole contributions of the respective form factor.
The strong coupling constant  $a_s$ in the $\overline{\rm MS}$ scheme obeys the scale evolution equation 
%--------------------------------------------------------------------------------------------------------
\begin{equation}
\frac{d a_s(\mu^2)}{d \ln \mu^2} = - \sum_{k=0}^\infty \beta_k a_s^{k+2}(\mu^2),
\end{equation}
%--------------------------------------------------------------------------------------------------------
where $\beta_{k}$ are the expansion coefficients of the QCD $\beta$--function \cite{Tarasov:1980au,Larin:1993tp,
vanRitbergen:1997va,Czakon:2004bu,Baikov:2016tgj,Herzog:2017ohr,Luthe:2017ttg}, which depend 
on the Casimir operators $C_A = N_C, C_F= (N^2_C-1)/(2 N_C)$ and $ T_{F} = 1/2$ in $SU(N_C)$ and the number of flavors $n_l$ 
up to three--loop order.
%--------------------------------------------------------------------------------------------------------
The functions $K_I$ incorporate the contributions from the heavy quark mass $m$ and are process independent, 
whereas, the functions $G_I$ are process dependent. The functions $G_I$ and $K_I$ obey the renormalization group 
equations 
%--------------------------------------------------------------------------------------------------------
\begin{equation} \label{eq:rge}
 \mu^2 \frac{d}{d \mu^2}  G_{I}\left(\frac{Q^2}{\mu^2},a_s,\ep\right) 
= - \lim_{m\rightarrow 0}  \mu^2 \frac{d}{d \mu^2} K_I\left(\frac{m^2}{\mu^2},a_s,\ep\right)
= A_q \Big( a_s (\mu^2) \Big),
\end{equation}
%--------------------------------------------------------------------------------------------------------
cf. also \cite{Mitov:2006xs}. Individually one obtains
%--------------------------------------------------------------------------------------------------------
\begin{align}
 K_{I} &=
 K_{I} \Big( a_s (m^2), 1, \ep \Big) - \int_{\frac{m^2}{\mu^2}}^{1}  \frac{d \lambda}{\lambda} A_q \Big( a_s 
(\lambda 
\mu^2) \Big),
\nonumber\\
 G_{I} &=
 G_{I} \Big( a_s (Q^2), 1, \ep \Big) + \int_{\frac{Q^2}{\mu^2}}^{1}  \frac{d \lambda}{\lambda} A_q \Big( a_s 
(\lambda 
\mu^2) \Big).
\end{align}
%--------------------------------------------------------------------------------------------------------
It is understood that all contributing functions obey series expansions of the kind
%--------------------------------------------------------------------------------------------------------
\begin{equation} 
{\cal A}_{I}(a_s) = \sum_{k=1}^\infty a_s^k {\cal A}_{I}^{(k)}.
\end{equation}
%--------------------------------------------------------------------------------------------------------
The coefficients $A_q^{(n)}$ are known up to three--loop order \cite{Moch:2004pa, Vogt:2004mw}.
The leading color and complete $n_l^2,~ n_l^3$ contributions of $A_q^{(4)}$ have been obtained
in \cite{Henn:2016men,Davies:2016jie,Lee:2016ixa,Lee:2017mip,Moch:2017uml,Moch:2018wjh} recently.
The finite functions $K_I$, $G_I$ and $A_q$ depend on $a_s(k^2)$
with the corresponding scales $k^2 = m^2, Q^2, \lambda \mu^2$.

To solve Eq.~(\ref{eq:kge}), the following expansion is performed
%--------------------------------------------------------------------------------------------------------
\begin{equation}  \label{eq:lnfexp}
 \ln \hat{F}_{I} \Big( a_s, \frac{Q^2}{\mu^2}, \frac{m^2}{\mu^2}, \ep \Big) 
= \sum_{n=1}^{\infty} {a}_s^n \Big( \frac{m^2}{\mu^2} \Big)^{-n \ep} {\cal F}_{I}^{(n)}
\left(\frac{Q^2}{m^2},\frac{m^2}{\mu^2},\ep\right).
\end{equation}
%--------------------------------------------------------------------------------------------------------
In the following we use $L \equiv \ln\left(Q^2/m^2\right)$ by setting $\mu^2=m^2$, for convenience, cf. also 
\cite{Ahmed:2017gyt}. The $\mu^2$--dependence can be easily recovered. The non-logarithmic contributions to the 
form factor are not contained in $\hat{F}_{I}$. They are obtained by matching
%--------------------------------------------------------------------------------------------------------
\begin{equation}  \label{eq:ffexp}
{F}_{I} \Big( a_s, \frac{Q^2}{\mu^2}, \frac{m^2}{\mu^2}, \ep \Big) 
= {\cal C}_I(a_s,\ep) \hat{F}_{I} \Big( a_s, \frac{Q^2}{\mu^2}, \frac{m^2}{\mu^2}, \ep \Big)
\end{equation}
%--------------------------------------------------------------------------------------------------------
to the complete form factors.

To have a more compact representation, the functions ${\cal F}_{I}^{(k)}$ in Eq.~(\ref{eq:lnfexp}) are 
presented
for the case of the unrenormalized coupling $\hat{a}_s$ here:
%--------------------------------------------------------------------------------------------------------
\begin{align}
{\cal F}_{I}^{(1)} &= 
\frac{1}{\varepsilon}   \bigg[ \bigg(  - \frac{1}{2} K_I^{(1)} - \frac{1}{2} G_I^{(1)} \bigg)
       + \frac{1}{2} A_I^{(1)} L \bigg]
       + \frac{1}{2} G_I^{(1)} L 
       - \frac{1}{4} A_I^{(1)} L^2
       + \varepsilon  \bigg[- \frac{1}{4} G_I^{(1)} L^2
\nonumber\\&       
       + \frac{1}{12} A_I^{(1)} L^3 \bigg]
       + \varepsilon^2  \bigg[ \frac{1}{12} G_I^{(1)} L^3
       - \frac{1}{48} A_I^{(1)} L^4 \bigg]
       + \varepsilon^3  \bigg[ - \frac{1}{48} G_I^{(1)} L^4 
       + \frac{1}{240} A_I^{(1)} L^5  \bigg] \,.
%%%%%%%%%%%%%%%%%%%%%%%%%%%%
\\
{\cal F}_{I}^{(2)} &= 
         \frac{1}{\varepsilon^2}   \bigg[  \bigg(  - \frac{1}{4} K_I^{(1)} \beta_0 - \frac{1}{4} G_I^{(1)} \beta_0 \bigg)
       + \frac{1}{4} A_I^{(1)} \beta_0 L \bigg]
       + \frac{1}{\varepsilon}   \bigg[ \bigg(  - \frac{1}{4} K_I^{(2)} - \frac{1}{4} G_I^{(2)} \bigg)
       + \bigg( \frac{1}{4} A_I^{(2)} 
\nonumber\\&       
+ \frac{1}{2} G_I^{(1)} \beta_0 \bigg) L
       - \frac{1}{4} A_I^{(1)} \beta_0 L^2 \bigg]
       + \frac{1}{2} G_I^{(2)} L
       + \bigg(  - \frac{1}{4} A_I^{(2)} - \frac{1}{2} G_I^{(1)} \beta_0 \bigg) L^2
\nonumber\\&       
       + \frac{1}{6} A_I^{(1)} \beta_0 L^3
       + \varepsilon  \bigg[  - \frac{1}{2} G_I^{(2)} L^2
       + \bigg( \frac{1}{6} A_I^{(2)} + \frac{1}{3} G_I^{(1)} \beta_0 \bigg) L^3
       - \frac{1}{12} A_I^{(1)} \beta_0 L^4 \bigg]
\nonumber\\ &
       + \varepsilon^2  \bigg[\frac{1}{3} G_I^{(2)} L^3
       + \bigg(  - \frac{1}{12} A_I^{(2)} - \frac{1}{6} G_I^{(1)} \beta_0 \bigg) L^4
       + \frac{1}{30} A_I^{(1)} \beta_0 L^5\bigg] \,.
%        
%%%%%%%%%%%%%%%%%%%%%%%%%%%%%%%
% 
\\
{\cal F}_{I}^{(3)} &= 
         \frac{1}{\varepsilon^3}   \bigg[ \bigg(  - \frac{1}{6} K_I^{(1)} \beta_0^2 - \frac{1}{6} G_I^{(1)} \beta_0^2 \bigg)
       + \frac{1}{6} A_I^{(1)} \beta_0^2 L \bigg]
       + \frac{1}{\varepsilon^2}   \bigg[ \bigg(  - \frac{1}{3} K_I^{(2)} \beta_0 - \frac{1}{12} K_I^{(1)} \beta_1 
- \frac{1}{3} G_I^{(2)} \beta_0 
\nonumber\\&       
- \frac{1}{12} G_I^{(1)} \beta_1 \bigg)
       + \bigg( \frac{1}{3} A_I^{(2)} \beta_0 + \frac{1}{12} A_I^{(1)} \beta_1 + \frac{1}{2} G_I^{(1)} \beta_0^2 \bigg) L
       - \frac{1}{4} A_I^{(1)} \beta_0^2 L^2 \bigg]
       + \frac{1}{\varepsilon}   \bigg[ - \frac{1}{6} (K_I^{(3)} +  G_I^{(3)}) 
\nonumber\\&       
       + \bigg( \frac{1}{6} A_I^{(3)} 
+ G_I^{(2)} \beta_0 + \frac{1}{4} G_I^{(1)} \beta_1 \bigg) L
       + \bigg(  - \frac{1}{2} A_I^{(2)} \beta_0 - \frac{1}{8} A_I^{(1)} \beta_1 - \frac{3}{4} G_I^{(1)} \beta_0^2 \bigg) L^2
       + \frac{1}{4} A_I^{(1)} \beta_0^2 L^3 \bigg]
\nonumber\\&       
       + \frac{1}{2} G_I^{(3)} L 
       + \bigg(  - \frac{1}{4} A_I^{(3)} 
- \frac{3}{2} G_I^{(2)} \beta_0 - \frac{3}{8} G_I^{(1)} \beta_1 \bigg) L^2
       + \bigg( \frac{1}{2} A_I^{(2)} \beta_0 + \frac{1}{8} A_I^{(1)} \beta_1 + \frac{3}{4} G_I^{(1)} \beta_0^2 \bigg) L^3
\nonumber\\ &
       - \frac{3}{16} A_I^{(1)} \beta_0^2  L^4
       + \varepsilon  \bigg[ - \frac{3}{4} G_I^{(3)} L^2
       + \bigg( \frac{1}{4} A_I^{(3)} + \frac{3}{2} G_I^{(2)} \beta_0 + \frac{3}{8} G_I^{(1)} \beta_1 \bigg) L^3
       + \bigg(- \frac{3}{8} A_I^{(2)} \beta_0 
\nonumber\\ &
- \frac{3}{32} A_I^{(1)} \beta_1 - \frac{9}{16} G_I^{(1)} \beta_0^2 \bigg) L^4
       + \frac{9}{80} A_I^{(1)} \beta_0^2 L^5 \bigg] \,.
\label{eq:anares3L}
%%%%%%%%%%%%%%%%%%%%%%%%%%%%%%%
% 
\\
{\cal F}_{I}^{(4)} &= 
         \frac{1}{\varepsilon^4}   \bigg[ \bigg(  - \frac{1}{8} K_I^{(1)} \beta_0^3 - \frac{1}{8} G_I^{(1)} \beta_0^3 \bigg)
       + \frac{1}{8} A_I^{(1)} \beta_0^3 L \bigg]
       + \frac{1}{\varepsilon^3}   \bigg[ \bigg(  - \frac{3}{8} K_I^{(2)} \beta_0^2 - \frac{1}{6} K_I^{(1)} \beta_0 \beta_1 
- \frac{3}{8} G_I^{(2)} \beta_0^2 
\nonumber\\&         
- \frac{1}{6} G_I^{(1)}
         \beta_0 \beta_1 \bigg)
       + \bigg( \frac{3}{8} A_I^{(2)} \beta_0^2 + \frac{1}{6} A_I^{(1)} \beta_0 \beta_1 + \frac{1}{2} G_I^{(1)} \beta_0^3 
\bigg) L
        - \frac{1}{4} A_I^{(1)} \beta_0^3  L^2 \bigg]
       + \frac{1}{\varepsilon^2}   \bigg[ \bigg(  - \frac{3}{8} K_I^{(3)} \beta_0 
\nonumber\\&       
- \frac{1}{8} K_I^{(2)} \beta_1 - \frac{1}{24} K_I^{(1)} \beta_2 - \frac{3}{8} G_I^{(3)} \beta_0 
       - \frac{1}{8} G_I^{(2)} \beta_1 - \frac{1}{24} G_I^{(1)} \beta_2 \bigg)
       +  \bigg( \frac{3}{8} A_I^{(3)} \beta_0 + \frac{1}{8} A_I^{(2)} \beta_1 
\nonumber\\ &
+ \frac{1}{24} A_I^{(1)} \beta_2 
+ \frac{3}{2} G_I^{(2)} \beta_0^2 + \frac{2}{3}
         G_I^{(1)} \beta_0 \beta_1 \bigg) L
       + \bigg(  - \frac{3}{4} A_I^{(2)} \beta_0^2 - \frac{1}{3} A_I^{(1)} \beta_0 \beta_1 
                    - G_I^{(1)} \beta_0^3 \bigg) L^2
\nonumber\\ &
       + \frac{1}{3} A_I^{(1)} \beta_0^3 L^3 \bigg]
       + \frac{1}{\varepsilon}   \bigg[ \bigg(  - \frac{1}{8} K_I^{(4)} - \frac{1}{8} G_I^{(4)} \bigg)
       +  \bigg( \frac{1}{8} A_I^{(4)} + \frac{3}{2} G_I^{(3)} \beta_0 + \frac{1}{2} G_I^{(2)} \beta_1 
+ \frac{1}{6} G_I^{(1)} \beta_2 \bigg) L
\nonumber\\ &
       + \bigg(  - \frac{3}{4} A_I^{(3)} \beta_0 
       - \frac{1}{4} A_I^{(2)} \beta_1 
- \frac{1}{12} A_I^{(1)} \beta_2 
- 3 G_I^{(2)} \beta_0^2
          - \frac{4}{3} G_I^{(1)} \beta_0 \beta_1 \bigg) L^2
       +  \bigg( A_I^{(2)} \beta_0^2 + \frac{4}{9} A_I^{(1)} \beta_0 \beta_1 
\nonumber\\ &
+ \frac{4}{3} G_I^{(1)} \beta_0^3 \bigg) L^3
       - \frac{1}{3} A_I^{(1)} \beta_0^3 L^4 \bigg]
       +  \frac{1}{2} G_I^{(4)}  L     
       + \bigg(  - \frac{1}{4} A_I^{(4)} - 3 G_I^{(3)} \beta_0 - G_I^{(2)} \beta_1 - \frac{1}{3} G_I^{(1)} \beta_2 \bigg) L^2
\nonumber\\ &
       + \bigg( A_I^{(3)} \beta_0 
+ \frac{1}{3} A_I^{(2)} \beta_1 + \frac{1}{9} A_I^{(1)} \beta_2 + 4 G_I^{(2)} \beta_0^2 
+ \frac{16}{9} G_I^{(1)} \beta_0
         \beta_1 \bigg) L^3
       + \bigg(  - A_I^{(2)} \beta_0^2 - \frac{4}{9} A_I^{(1)} \beta_0 \beta_1 
\nonumber\\ &
- \frac{4}{3} G_I^{(1)} \beta_0^3 \bigg) L^4
       + \frac{4}{15} A_I^{(1)} \beta_0^3 L^5 \,.
\label{eq:anares4L}
\end{align}
%----------------------------------------------------------------------------------------------------------
The corresponding expression for the running coupling $a_s$ is obtained by replacing 
%----------------------------------------------------------------------------------------------------------
\begin{eqnarray}
\hat{a}_s = 1 - a_s  \frac{\beta_0}{\ep}
              + a_s^2 \left(\frac{\beta_0^2}{\ep^2} - \frac{\beta_1}{2 \ep} \right)
              + a_s^3 \left(-\frac{\beta_0^3}{\ep^3} +\frac{7}{6\ep^2} \beta_0 \beta_1 - 
\frac{\beta_2}{3 \ep} \right) + {\cal O}(a_s^4)
\end{eqnarray}
%----------------------------------------------------------------------------------------------------------
in Eqs.~(\ref{eq:lnfexp},\ref{eq:ffexp}).

In massless scenarios, the soft ($f_i$) and collinear ($B_i$) anomalous dimensions of massless QCD 
govern the infrared structures, where $i \equiv q,g$ label quarks and gluons, respectively. 
In the case of two--parton amplitudes, e.g. the massless form factors,
$\gamma_i = B_i + \frac{f_i}{2}$ is obtained from the leading pole $1/\ep$ and terms from lower orders.
It is imperative that for the case of heavy quarks $\gamma_q$, along with similar contributions 
from the heavy quark anomalous dimension, will form the leading singularity. On the other hand, 
following soft-collinear effective theory, it was shown in \cite{Becher:2009kw} that the
two--loop anomalous dimension matrix for two heavy parton correlations, which controls the singular structure,
contains a heavy quark anomalous dimension $\gamma_Q$. Hence it is suggestive to form the 
following structure for the finite function 
%--------------------------------------------------------------------------------------------------------
\begin{align}
 K_I^{(n)} = -2 \left(\gamma_{q}^{(n)} + \gamma_{Q}^{(n)} - \gamma_{I}^{(n-1)}\right),
\end{align}
%--------------------------------------------------------------------------------------------------------
where $\gamma_q = B_q + \frac{f_q}{2}$ and $B_q$ and $f_q$ are the collinear and soft anomalous dimension 
for massless quarks. They are known up to three--loop order \cite{Moch:2004pa,Vogt:2004mw,Ravindran:2004mb}. 
Recently $\gamma_q^{(4)}$ in the color--planar limit has been calculated in \cite{Henn:2016men,Lee:2016ixa,Lee:2017mip} from 
the contributions to the four--loop massless form factor in addition to the $n_l^2$ and $n_l^3$ contributions.
The anomalous dimension $\gamma_I^{(n-1)}$ is  arising due to renormalization of the current. It vanishes
for conserved currents, as the vector current. For the scalar current, the Yukawa coupling introduces 
the mass renormalization and hence $\gamma_S^{(n-1)}=\gamma_m^{(n-1)}$,
being known up to four--loop order \cite{Broadhurst:1991fy,Vermaseren:1997fq,Gracey:2000am,
Melnikov:2000zc,Marquard:2007uj,Luthe:2016xec,Baikov:2017ujl}. Furthermore, the renormalization constants
$Z_{m, {\rm OS}}$  and $Z_{2,{\rm OS}}$ are given in  
\cite{Broadhurst:1991fy, Melnikov:2000zc,Marquard:2007uj,Marquard:2015qpa,Marquard:2016dcn},
and
\cite{Marquard:2015qpa,Marquard:2016dcn,Marquard:2018rwx}, respectively.

The heavy quark anomalous dimension $\gamma_Q$ is connected with the soft Wilson-line operator
and was known up to two-loop \cite{Becher:2009kw, Neubert:2004dd,Gardi:2004ia}. Also, it constitutes the 
non-logarithmic contribution of the massive cusp anomalous dimension in the asymptotic limit. Thus, we can 
obtain $\gamma_Q$ up to three-loop order from the known three-loop result \cite{Grozin:2014hna} 
%--------------------------------------------------------------------------------------------------------
\begin{align}
 \gamma_Q^{(0)} &= - 2 C_F \,,
 \nonumber\\
 \gamma_Q^{(1)} &= 
 C_F C_A  \left(
-\frac{98}{9}
+4 \zeta_2
-4 \zeta_3
\right)
+ C_F T_F n_l  \bigg(
\frac{40}{9}
\bigg) \,,
 \nonumber\\
\gamma_Q^{(2)} &= 
 C_F C_A^2  \bigg(
-\frac{343}{9}
+\frac{608}{9} \zeta_2
-\frac{88}{5} \zeta_2^2
-\frac{740}{9} \zeta_3
-8 \zeta_2 \zeta_3
+36 \zeta_5
\bigg)
% \nonumber\\
% &
+ C_F C_A T_F n_l  \bigg(
\frac{356}{27}
-\frac{160}{9} \zeta_2
\nonumber\\&
+\frac{496}{9} \zeta_3
\bigg)
+ C_F^2 T_F n_l  \bigg(
\frac{110}{3}
-32 \zeta_3
\bigg)
+ C_F T_F^2 n_l^2  \bigg(
\frac{32}{27}
\bigg) \,.
\end{align}
%--------------------------------------------------------------------------------------------------------
Here $\zeta_k,~k \geq 2, k \in \mathbb{N}$ are the values of the Riemann $\zeta$-function at integer argument.
Note that, as mentioned in \cite{Becher:2009kw}, the non-abelian exponentiation theorem constrains the color 
structures of $\gamma_Q$.
Hence, up to ${\cal O}(\alpha_s^3)$, the corresponding anomalous
dimension for a massive color--octet boson ($G$) is given by $\gamma_G = \frac{C_A}{C_F} \gamma_Q$.
The other finite function $G_I^{(n)}$ only contains the information about the process and depends on $Q^2$
and not the heavy quark mass $m$. 
It is similar to the massless case of the Drell-Yan form factor
%--------------------------------------------------------------------------------------------------------
\begin{equation} \label{eq:gin}
 G_I^{(n)} = 2 (B_q^{(n)} - \gamma_I^{(n-1)}) + f_q^{(n)} + C_I^{(n)} + \sum_{k=1}^{\infty} \ep^k g_{I}^{n,k}. 
\end{equation}
%--------------------------------------------------------------------------------------------------------
Up to four--loop order, $C_I^{(n)}$ are given by
%--------------------------------------------------------------------------------------------------------
\begin{align}
 C_I^{(1)} &= 0 \,,
  \nonumber\\
 C_I^{(2)} &= \beta_0 g_{I}^{1,1} \,,
  \nonumber\\
 C_I^{(3)} &= \beta_1 g_{I}^{1,1} + \beta_0 ( g_{I}^{2,1} - \beta_0 g_{I}^{1,2} ) \,,
 \nonumber\\
 C_I^{(4)} &= \beta_2 g_{I}^{1,1} + \beta_1 ( g_{I}^{2,1} - 2 \beta_0 g_{I}^{1,2} ) 
        + \beta_0 ( g_{I}^{3,1} - \beta_0 g_{I}^{2,2} 
        + \beta_0^2 g_{I}^{1,3} ) \,.
\end{align}
%--------------------------------------------------------------------------------------------------------
The components $g_{I}^{n,k}$ have to be extracted from explicit result in this limit.
Given the structural similarities, the functions $g_{I}^{n,k}$ are the same in both the massive and the 
massless case.
%--------------------------------------------------------------------------------------------------------
\section{Phenomenological Results} \label{sec:result}
%--------------------------------------------------------------------------------------------------------

\vspace*{1mm}
\noindent 
Eqs.~(\ref{eq:anares3L}) and (\ref{eq:anares4L}), along with (\ref{eq:lnfexp}) and (\ref{eq:ffexp}), provide the 
three-- and four--loop heavy quark form factors in the asymptotic limit. To obtain the vector and scalar form 
factors, we need ${\cal C}_I^{(3)}$ and ${\cal C}_I^{(4)}$, along with all the anomalous dimensions up to 
three and four loops and $g_I^{1,k}, g_I^{2,k}, g_I^{3,k}, \cC_I^{1,k}$ and $\cC_I^{2,k}$ to sufficient order 
$k$ in $\ep$.
As noted earlier, the coefficients $g_I^{n,k}$ can be obtained from the massless results, 
cf.~\cite{Ahmed:2014cla,Ahmed:2014cha}.
The $\varepsilon^0$--part of $G_I^{(4)}$, as noted in Eq.~(\ref{eq:gin}), can also be obtained from the  massless quark 
form factor.
Below, we present $G_V^{(4)}$ and $G_S^{(4)}$. Note that $G_S^{(4)}$ constitutes the single pole in 
$\varepsilon$ of the Higgs--quark massless form factor.
%--------------------------------------------------------------------------------------------------------
\begin{align}
G_V^4 &=
N_C^4 \bigg(
 \frac{1913092765}{139968}
+\frac{571537}{243} \zeta_2
-\frac{2308}{5} \zeta_2^2
+\frac{12350}{63} \zeta_2^3
-\frac{169547}{18} \zeta_3
+\frac{1622}{3} \zeta_3^2
+\frac{8480}{9} \zeta_3 \zeta_2
\nonumber\\&
-\frac{656}{5} \zeta_3 \zeta_2^2
+\frac{10100}{3} \zeta_5
-288 \zeta_5 \zeta_2
-1410 \zeta_7
\bigg)
+ N_C^3 n_l T_F \bigg(
-\frac{164021633}{11664}
-\frac{1903985}{486} \zeta_2
+\frac{7547}{15} \zeta_2^2
\nonumber\\&
-\frac{29432}{315} \zeta_2^3
+\frac{406621}{81} \zeta_3
+\frac{32}{3} \zeta_3^2
-\frac{2432}{9} \zeta_3 \zeta_2
+\frac{1340}{9} \zeta_5
\bigg)
+ C_F^2 n_l^2 T_F^2 \bigg(
 \frac{693446}{243}
-\frac{3232}{3} \zeta_2
\nonumber\\ &
-\frac{21376}{45} \zeta_2^2
+\frac{65632}{27} \zeta_3
-\frac{512}{3} \zeta_3 \zeta_2
-\frac{832}{3} \zeta_5
\bigg)
+ C_F C_A n_l^2 T_F^2 \bigg(
 \frac{8996449}{1458}
+\frac{919544}{243} \zeta_2
+\frac{2464}{15} \zeta_2^2
\nonumber\\&
-\frac{172832}{81} \zeta_3
+\frac{1216}{9} \zeta_3 \zeta_2
-\frac{4928}{9} \zeta_5
\bigg)
+ C_F n_l^3 T_F^3  \bigg(
-\frac{1117984}{2187}
-\frac{25984}{81} \zeta _2
-\frac{1664}{45} \zeta _2^2
-\frac{640}{81} \zeta_3
\bigg)
\nonumber\\&
+{\cal O}(\ep) \,.
% 
%%%%%%%%%%%%%%%%%%%%%%%%%%%%%%%%% GS4
\\
G_S^4 &=
N_C^4 \bigg(
-\frac{11003395}{4374}
+\frac{93937}{972} \zeta_2
-290 \zeta_2^2
+\frac{12350}{63} \zeta_2^3
-\frac{36592}{9} \zeta_3
+\frac{1622}{3} \zeta_3^2
+\frac{8480}{9} \zeta_3 \zeta_2
\nonumber\\ &
-\frac{656}{5} \zeta_3 \zeta_2^2
+\frac{8780}{3} \zeta_5
-288 \zeta_5 \zeta_2
-1410 \zeta_7
\bigg)
+ N_C^3 n_l T_F \bigg(
 \frac{14211811}{11664}
-\frac{497663}{486} \zeta_2
+\frac{5027}{15} \zeta_2^2
\nonumber\\ &
-\frac{29432}{315} \zeta_2^3
+\frac{199909}{81} \zeta_3
+\frac{32}{3} \zeta_3^2
-\frac{2432}{9} \zeta_3 \zeta_2
\nonumber\\&
+\frac{620}{9} \zeta_5
\bigg)
+ C_F^2 n_l^2 T_F^2 \bigg(
 \frac{201110}{243}
-\frac{2944}{9} \zeta_2
-\frac{16192}{45} \zeta_2^2
+\frac{67360}{27} \zeta_3
-\frac{512}{3} \zeta_3 \zeta_2
-\frac{832}{3} \zeta_5
\bigg)
\nonumber\\&
+ C_F C_A n_l^2 T_F^2 \bigg(
-\frac{823055}{1458}
+\frac{293144}{243} \zeta_2
+\frac{2752}{15} \zeta_2^2
-\frac{136544}{81} \zeta_3
+\frac{1216}{9} \zeta_3 \zeta_2
-\frac{4928}{9} \zeta_5
\bigg)
\nonumber\\&
+ C_F n_l^3 T_F^3  \bigg(
-\frac{41728}{2187}
-\frac{6400}{81} \zeta_2
-\frac{1664}{45} \zeta_2^2
-\frac{640}{81} \zeta_3
\bigg) 
+{\cal O}(\ep) \,.
\end{align}
%--------------------------------------------------------------------------------------------------------
Due to the recent two--loop calculations in Ref.~\cite{Ablinger:2017hst} we can extract the coefficients in
%--------------------------------------------------------------------------------------------------------
\begin{align}
{\cal C}^{(i)} &= \sum_{k = 0}^\infty \ep^k {\cal C}^{i,k}  
\end{align}
%--------------------------------------------------------------------------------------------------------
up to the required order. They read
%--------------------------------------------------------------------------------------------------------
\begin{align}
\cC_{V}^{1,0} &= C_F \bigg(
4
+\zeta_2
 \bigg) \,,
  \nonumber\\
%%%%%%%%%%%%%%%  
%-----
\cC_{V}^{1,1} &= C_F \bigg(
8
+\frac{1}{2} \zeta_2
-\frac{2}{3} \zeta_3
 \bigg) \,,
  \nonumber\\
%-----
 \cC_{V}^{1,2} &= C_F \bigg(
16
+2 \zeta_2
+\frac{9}{20} \zeta_2^2
-\frac{1}{3} \zeta_3
 \bigg)
 \nonumber\\
%-----
 \cC_{V}^{1,3} &= C_F \bigg(
32
+4 \zeta_2
+\frac{9}{40} \zeta_2^2
-\frac{4}{3} \zeta_3
-\frac{1}{3} \zeta_2 \zeta_3
-\frac{2}{5} \zeta_5
 \bigg) \,,
 \nonumber\\
%-----
\cC_{V}^{2,0} &= C_F^2 \bigg(
\frac{241}{8}
+26 \zeta_2
-48 \ln (2) \zeta_2
-\frac{163}{10} \zeta_2^2
-6 \zeta_3
 \bigg)
% \nonumber\\& 
+ C_F C_A \bigg(
\frac{12877}{648}
+\frac{323}{18} \zeta_2
+24 \ln (2) \zeta_2
\nonumber\\&
-\frac{47}{5} \zeta_2^2
% \nonumber\\&
+\frac{89}{9} \zeta_3
\bigg)
+ C_F T_F n_l  \bigg(
-\frac{1541}{162}
-\frac{74}{9} \zeta_2
-\frac{52}{9} \zeta_3
\bigg) \,,
 \nonumber\\
%  
%%%%%%%%%%%%%%
\cC_{V}^{2,1} &= C_F^2 \bigg(
-\frac{557}{16}
+8 c_1
+\frac{399}{4} \zeta_2
-24 \ln (2) \zeta_2
-\frac{1363}{10} \zeta_2^2
+\frac{286}{3} \zeta_3
-\frac{146}{3} \zeta_2 \zeta_3
-6 \zeta_5
 \bigg)
\nonumber\\ &
+ C_F C_A \bigg(
\frac{629821}{3888}
-4 c_1
+\frac{7805}{108} \zeta_2
+12 \ln (2) \zeta_2
+\frac{209}{2} \zeta_2^2
-\frac{1162}{27} \zeta_3
+\frac{83}{3} \zeta_2 \zeta_3
-208 \zeta_5
 \bigg)
\nonumber\\ &
+ C_F n_l T_F \bigg(
-\frac{46205}{972}
-\frac{673}{27} \zeta_2
-\frac{98}{5} \zeta_2^2
-\frac{364}{27} \zeta_3
\bigg) \,,
%  
%%%%%%%%%%%%%%%
\nonumber\\
\cC_{V}^{2,2} &=
C_F^2 \bigg(
\frac{1817}{32}
+4 c_1
+\frac{48}{5} c_2
+8 c_1 \zeta_2
+\frac{5165}{8} \zeta_2
-576 \zeta_2 \ln (2)
+\frac{393}{8} \zeta_2^2
-\frac{37137}{140} \zeta_2^3
-144 \zeta_2^2 \ln^2(2)
\nonumber\\&
+\frac{1889}{6} \zeta_3
-\frac{799}{9} \zeta_3^2
-\frac{494}{3} \zeta_3 \zeta_2
+168 \zeta_3 \zeta_2 \ln (2)
-1431 \zeta_5
\bigg)
+ C_F C_A  \bigg(
\frac{10899301}{23328}
-2 c_1
-4 c_1 \zeta_2
\nonumber\\&
-\frac{24}{5} c_2
+\frac{53045}{648} \zeta_2
+\frac{2513}{24} \zeta_2^2
-\frac{9839}{210} \zeta_2^3
+288 \zeta_2 \ln (2)
+72 \zeta_2^2 \ln^2(2)
-\frac{6517}{162} \zeta_3
+\frac{13}{2} \zeta_3^2
\nonumber\\&
+\frac{1387}{9} \zeta_3 \zeta_2
-84 \zeta_3 \zeta_2 \ln (2)
+\frac{23077}{30} \zeta_5   
\bigg)
+ C_F n_l T_F  \bigg(
-\frac{1063589}{5832}
-\frac{15481}{162} \zeta_2
-\frac{1607}{30} \zeta_2^2
\nonumber\\ &
-\frac{3422}{81} \zeta_3
-\frac{224}{9} \zeta_3 \zeta_2
-\frac{1444}{15} \zeta_5
\bigg) 
\nonumber\\
%%%%%%%%%%%%
{\cal C}_V^{3,0} &= 
2 C_F^2 T_F n_l \Biggl(
        -
        \frac{78427}{972}
        -\frac{32}{9} c_1
        -\frac{38267}{324} \zeta_2
        +\frac{224}{3} \ln(2) \zeta_2
        +\frac{17873}{135} \zeta_2^2
        -\frac{5264}{81} \zeta_3
        -\frac{154}{9} \zeta_2 \zeta_3
\nonumber\\ &
        +\frac{596}{9} \zeta_5
\Biggr)
+ N_C^3 \Biggl(
        \frac{10907077}{52488}
        +\frac{15763067}{46656} \zeta_2
        -\frac{116957}{1080} \zeta_2^2
        +\frac{145051}{3024} \zeta_2^3
        -\frac{15743}{486} \zeta_3
\nonumber\\ &
        +\frac{533}{18} \zeta_2 \zeta_3
        +\frac{298}{9} \zeta_3^2
        -357 \zeta_5
\Biggr)
+2  C_F C_A T_F n_l \Biggl(
                -\frac{533996}{6561}
                +\frac{16}{9} c_1
                -\frac{102019}{729} \zeta_2
                -\frac{112}{3} \ln(2) \zeta_2
\nonumber\\ &
                -\frac{10942}{135} \zeta_2^2
                -\frac{2696}{81} \zeta_3
                +\frac{302}{3} \zeta_5
        \Biggr)
        + 4 C_F T_F^2 n_l^2 \Biggl(
                \frac{58883}{13122}
                +\frac{580}{81} \zeta_2
                +\frac{884}{135} \zeta_2^2
                +\frac{2144}{243} \zeta_3
        \Biggr)
\\
\cC_{S}^{1,0} &= C_F \zeta_2
\nonumber\\ 
\cC_{S}^{1,1} &= C_F \bigg(
-\zeta_2
-\frac{2}{3} \zeta_3
 \bigg) 
\nonumber\\  
\cC_{S}^{1,2} &= C_F \bigg(
\frac{9}{20} \zeta_2^2
+\frac{2}{3} \zeta_3
 \bigg) 
\nonumber\\
%%%%%%%%%%%%%%%%%
\cC_{S}^{1,3} &= C_F \bigg(
-\frac{9}{20} \zeta_2^2
-\frac{1}{3} \zeta_2 \zeta_3
-\frac{2}{5} \zeta_5
 \bigg) \,,
\nonumber\\
% 
%%%%%%%%%%%%%%%
\cC_{S}^{2,0} &= C_F^2 \bigg(
15
-8 \zeta_2
-\frac{163}{10} \zeta_2^2
-18 \zeta_3
\bigg)
+ C_F C_A \bigg(
-\frac{2140}{81}
+\frac{467}{18} \zeta_2
-\frac{47}{5} \zeta_2^2
+\frac{143}{9} \zeta_3
 \bigg)
\nonumber\\ &
+ C_F n_l T_F \bigg(
\frac{188}{81}
-\frac{2}{9} \zeta_2
-\frac{52}{9} \zeta_3
\bigg)
 \nonumber\\
 \cC_{S}^{2,1} &= C_F^2 \bigg(
-\frac{209}{2}
-76 \zeta_2
+264 \ln (2) \zeta_2
-37 \zeta_2^2
-34 \zeta_3
-\frac{146}{3} \zeta_2 \zeta_3
-6 \zeta_5
 \bigg)
+ C_F C_A \bigg(
-\frac{8155}{486}
\nonumber\\ &
+\frac{2561}{27} \zeta_2
-132 \ln (2) \zeta_2
+\frac{541}{10} \zeta_2^2
+\frac{242}{27} \zeta_3
+\frac{83}{3} \zeta_2 \zeta_3
-208 \zeta_5
 \bigg)
+ C_F n_l T_F \bigg(
\frac{214}{243}
+\frac{200}{27} \zeta_2
\nonumber\\ &
-\frac{98}{5} \zeta_2^2
+\frac{68}{27} \zeta_3
\bigg) \,,
% 
%%%%%%%%%
\nonumber\\
\cC_{S}^{2,2} &=
C_F^2 \bigg(
-\frac{779}{4}
-44 c_1
+8 c_1 \zeta _2
-17 \zeta _2
+480 \zeta _2 \ln (2)
+\frac{857}{2} \zeta_2^2
-\frac{37137}{140} \zeta_2^3
-144 \zeta_2^2 \ln^2 (2)
\nonumber\\ &
-272 \zeta_3
-\frac{799}{9} \zeta _3^2
-\frac{92}{3} \zeta_3 \zeta_2
+168 \zeta_3 \zeta_2 \ln (2)
-213 \zeta_5
\Bigg)
+ C_F C_A \bigg(
-\frac{415831}{2916}
+22 c_1
-4 c_1 \zeta_2
\nonumber\\ &
+\frac{11818}{81} \zeta_2
-240 \zeta_2 \ln (2)
-\frac{2098}{15} \zeta_2^2
-\frac{9839}{210} \zeta_2^3
+72 \zeta_2^2 \ln^2 (2)
+\frac{14422}{81} \zeta_3
+\frac{13}{2} \zeta_3^2
+\frac{793}{9} \zeta_3 \zeta_2
\nonumber\\&
-84 \zeta_3 \zeta_2 \ln (2)
+\frac{4807}{30} \zeta_5
\bigg)
+ C_F n_l T_F \bigg(
-\frac{9049}{729}
+\frac{940}{81} \zeta_2
+\frac{92}{15} \zeta_2^2
+\frac{880}{81} \zeta_3
-\frac{224}{9} \zeta_3 \zeta_2
\nonumber\\ &
-\frac{1444}{15} \zeta_5
\bigg) 
\nonumber\\ 
{\cal C}_S^{3,0} &= 
   N_C^3 \Biggl(
-\frac{10407317}{52488} 
+ \frac{324905}{2916} \zeta_2 
- \frac{77213}{1080} \zeta_2^2 
+ \frac{145051}{3024} \zeta_2^3 
- \frac{28865}{486} \zeta_3
+ \frac{1181}{18} \zeta_2 \zeta_3 + \frac{298}{9} \zeta_3^2 
\nonumber\\ &
- 372 \zeta_5 \Biggr)
+ 
  2  C_F^2 T_F n_l \Biggl(
\frac{32351}{972} + \frac{9545}{162} \zeta_2 
- 160 \ln(2) \zeta_2 + \frac{1969}{27} \zeta_2^2 
+ \frac{1612}{81} \zeta_3 
- \frac{154}{9} \zeta_2 \zeta_3 
\nonumber\\ &
+ \frac{596}{9} \zeta_5\Biggr)
          + 
    2 C_F C_A T_F n_l \Biggl(
\frac{842609}{13122} - \frac{45319}{729} \zeta_2
 +  80 \ln(2) \zeta_2 - \frac{9574}{135} \zeta_2^2 
- \frac{1094}{81} \zeta_3 + \frac{302}{3} \zeta_5 \Biggr)
\nonumber\\ &
          + 
  4  C_F T_F^2 n_l^2 \Biggl(
-\frac{2324}{6561} 
- \frac{356}{81} \zeta_2 + \frac{884}{135} \zeta_2^2
+ \frac{632}{243}  \zeta_3 \Biggr)
\,.
\end{align}
%--------------------------------------------------------------------------------------------------------
The functions $\cC_{V}^{3,0}$ and $\cC_{S}^{3,0}$ have been obtained from the results in \cite{Henn:2016tyf, 
Ablinger:2018yae, FORMF3,Lee:2018rgs}. Here ${\cal C}_V$  and  ${\cal C}_S$ have been calculated individually.
One may check that they are related by
%--------------------------------------------------------------------------------------------------------
\begin{equation}
{\cal C}_S = \frac{Z_{m,\rm OS}}{Z_{m, \overline{\rm MS}}} {\cal C}_V.
\end{equation}
%--------------------------------------------------------------------------------------------------------

With all the available components, we can now present all the logarithmic contributions 
to the finite part for three--loop vector and scalar form factors in the asymptotic limit:
%--------------------------------------------------------------------------------------------------------
\begin{align}
F_{V}^{(3)} &=
\frac{1}{\ep^3} \bigg[
%%%%%%%
-C_F^3 \frac{4}{3} (1-L)^3
%%%%%%%
- C_F^2 C_A \frac{22}{3} (1-L)^2
- C_F C_A^2 \frac{242}{27} (1-L)
+ C_F C_A n_l T_F 
\frac{176}{27} (1-L)
\nonumber\\&
+ C_F^2 n_l T_F \frac{8}{3} (1-L)^2
- C_F n_l^2 T_F^2 
\frac{32}{27} (1-L)
\bigg]
%%%%%%%%%%%%%%%%%%%%%%%%%%
% \nonumber\\&
+
\frac{1}{\ep^2} \bigg[
%%%%%%%
C_F^3 \bigg\{
-8
+4 \zeta_2
+\Big(
        22
        -8 \zeta_2
\Big) L
\nonumber\\ &
+\Big(
        -22
        +4 \zeta_2
\Big) L^2
+10 L^3
-2 L^4 
\bigg\}
+ C_F^2 C_A \bigg\{
-\frac{34}{9}
+\frac{10}{3} \zeta_2
+4 \zeta_3
+\bigg(
        -\frac{1}{9}
        +\frac{2 \zeta_2}{3}
        -4 \zeta_3
\bigg) L
\nonumber\\&
+\bigg(
        \frac{2}{9}
        -4 \zeta_2
\bigg) L^2
+\frac{11 L^3}{3} 
\bigg\}
+ C_F C_A^2 \bigg\{
\frac{1690}{81}
-\frac{44}{9} \zeta_2
+\frac{44}{9} \zeta_3
+\bigg(
        -\frac{2086}{81}
        +\frac{44 \zeta_2}{9}
\bigg) L
\bigg\}
\nonumber\\ &
+ C_F C_A n_l T_F \bigg\{
-\frac{1192}{81}
+\frac{16}{9} \zeta_2
-\frac{16}{9} \zeta_3
+\bigg(
        \frac{1336}{81}
        -\frac{16 \zeta_2}{9}
\bigg) L 
\bigg\}
\\&
+ C_F^2 n_l T_F \bigg\{
-\frac{16}{9}
-\frac{8}{3} \zeta_2
+\bigg(
        \frac{20}{9}
        +\frac{8 \zeta_2}{3}
\bigg) L
+\frac{8 L^2}{9}
-\frac{4 L^3}{3} 
\bigg\}
+ C_F n_l^2 T_F^2 
   \frac{160}{81}(1-L)
\bigg]
%%%%%%%%%%%%%%%%%%%%%%%%%%
\nonumber\\ &
+
\frac{1}{\ep} \bigg[
%%%%%%%
C_F^3 \bigg\{
-76
-82 \zeta_2
+96 \ln (2) \zeta_2
+\frac{236}{5} \zeta_2^2
+80 \zeta_3
+\Big(
        129
        +88 \zeta_2
        -96 \ln (2) \zeta_2
        -\frac{236}{5} \zeta_2^2
\nonumber\\ &
        -136 \zeta_3
\Big) L
+\Big(
        -89
        +56 \zeta_3
\Big)L^2
+\bigg(
        \frac{137}{3}
        -6 \zeta_2
\bigg) L^3
-\frac{34 L^4}{3}
+\frac{5 L^5}{3} 
\bigg\}
+ C_F^2 C_A \bigg\{
\frac{2986}{27}
-10 \zeta_2
\nonumber\\ &
-48 \ln (2) \zeta_2
+\frac{26}{5} \zeta_2^2
-\frac{200}{3} \zeta_3
-4 \zeta_2 \zeta_3
+\bigg(
        -\frac{5396}{27}
        +\frac{5}{3} \zeta_2
        +48 \ln (2) \zeta_2
        -\frac{26}{5} \zeta_2^2
        +\frac{362}{3} \zeta_3
\bigg) L
\nonumber\\ &
+\bigg(
        \frac{6107}{54}
        +\frac{19 \zeta_2}{3}
        -50 \zeta_3
\bigg)L^2
+\bigg(
        -\frac{523}{18}
        +6 \zeta_2
\bigg) L^3
+\frac{11 L^4}{9} 
\bigg\}
+ C_F C_A^2 \bigg\{
-\frac{343}{27}
+\frac{608}{27} \zeta_2
\nonumber\\ &
-\frac{88}{15} \zeta_2^2
-\frac{740}{27} \zeta_3
-\frac{8}{3} \zeta_2 \zeta_3
+12 \zeta_5
+\bigg(
        \frac{245}{9}
        -\frac{536}{27} \zeta_2
        +\frac{88}{15} \zeta_2^2
        +\frac{44}{9} \zeta_3
\bigg) L
\bigg\}
+ C_F C_A n_l T_F \bigg\{
\frac{356}{81}
\nonumber\\ &
-\frac{160}{27} \zeta_2
+\frac{496}{27} \zeta_3
+\bigg(
        -\frac{836}{81}
        +\frac{160 \zeta_2}{27}
        -\frac{112 \zeta_3}{9}
\bigg) L 
\bigg\}
+ C_F^2 n_l T_F \bigg\{
-\frac{470}{27}
+8 \zeta_2
-\frac{16}{3} \zeta_3
\nonumber\\ &
+\bigg(
        \frac{1198}{27}
        -\frac{4 \zeta_2}{3}
        +\frac{16 \zeta_3}{3}
\bigg) L
+\bigg(
        -\frac{962}{27}
        -\frac{20 \zeta_2}{3}
\bigg) L^2
+\frac{82 L^3}{9}
-\frac{4 L^4}{9} 
\bigg\}
+ C_F n_l^2 T_F^2
\frac{32}{81}(1-L)
\bigg]
\nonumber\\ &
%%%%%%%%%%%%%%%%%%%%%%%%%%
+
 \bigg[
%%%%%%%
X_{V,3}^{0,0}
+ N_C^3 \bigg\{
-\frac{554267}{2916} 
+\frac{23773 \zeta_2}{216}
-\frac{1727 \zeta_2^2}{30}
+\frac{8156 \zeta_2^3}{315}
+\frac{33197 \zeta_3}{162}
-\frac{16 \zeta_3^2}{3}
+\frac{113 \zeta_3 \zeta_2}{3}
\nonumber\\&
-\frac{875 \zeta_5}{3}
\bigg\}
+ C_F^3  \bigg\{
\Big(
        250
        +16 c_1        
        +585 \zeta_2
        -192 \ln (2) \zeta_2
        -\frac{1942}{5} \zeta_2^2
        -520 \zeta_3
        -8 \zeta_2 \zeta_3
        -276 \zeta_5
\Big) L
\nonumber\\ &
-\Big(
        332
        +123 \zeta_2
        -48 \ln (2) \zeta_2        
        -\frac{302}{5} \zeta_2^2
        -340 \zeta_3
\Big) L^2
+ \bigg(
        \frac{494}{3}
        +\frac{17 \zeta_2}{3}
        -\frac{268 \zeta_3}{3}
\bigg) L^3
-\bigg(\frac{148}{3}
\nonumber\\&        
        -\frac{16 \zeta_2}{3}
\bigg) L^4
+\frac{17 L^5}{2}
-L^6
\bigg\}
% 
%
%%%%%%%%%%%%%%%%%%%%%%%%%%%%
%
+ C_F^2 C_A   \bigg\{
\bigg(
        -\frac{178337}{162}
        -8 c_1
        -\frac{5959}{18} \zeta_2
        +96 \ln (2) \zeta_2
        +\frac{4279}{15} \zeta_2^2
\nonumber\\&        
        +\frac{3706}{3} \zeta_3
        -46 \zeta_2 \zeta_3
        -194 \zeta_5
\bigg) L
+\bigg(
        \frac{64625}{81}
        +\frac{266}{3} \zeta_2
        -24 \ln (2) \zeta_2
        -\frac{143}{5} \zeta_2^2
        -\frac{3743}{9} \zeta_3
\bigg) L^2
\nonumber\\ &
+ \bigg(
        -\frac{6260}{27}
        -\frac{97 \zeta_2}{18}
        +\frac{232 \zeta_3}{3}
\bigg) L^3
+\bigg(
        \frac{4289}{108}
        -\frac{16 \zeta_2}{3}
\bigg) L^4
-\frac{11 L^5}{4}
\bigg\}
% %
% 
+ C_F C_A^2   \bigg\{
\bigg(
        \frac{1045955}{1458}
\nonumber\\ &
        +\frac{17366}{81} \zeta_2
        -\frac{94}{3} \zeta_2^2
        -\frac{17464}{27} \zeta_3
        +\frac{88}{3} \zeta_2 \zeta_3
        +136 \zeta_5
\bigg) L
+ \bigg(
        -\frac{18682}{81}
        +\frac{26}{9} \zeta_2        
        -\frac{44}{5} \zeta_2^2
        +88 \zeta_3
\bigg) L^2
\nonumber\\ &
+ \bigg(
        \frac{2869}{81}
        -\frac{44 \zeta_2}{9}
\bigg) L^3
-\frac{121 L^4}{54}
\bigg\}
+ C_F C_A n_l T_F \bigg\{
\frac{259150}{729}
+\frac{32 c_1}{9}
+\frac{3008 \zeta_2}{81}
-\frac{7288 \zeta_2^2}{45}
\nonumber\\ &
-\frac{224}{3} \zeta_2 \ln (2)
-\frac{31120 \zeta_3}{81}
+\frac{8 \zeta_3 \zeta_2}{3}
+\frac{596 \zeta_5}{3}
+ \bigg(
        -\frac{309838}{729}
        -\frac{11728}{81} \zeta_2
        +\frac{88}{15} \zeta_2^2
        +\frac{1448}{9} \zeta_3
\bigg) L
\nonumber\\ &
+\bigg(
        \frac{11752}{81}
        +\frac{32 \zeta_2}{3}
        -16 \zeta_3
\bigg) L^2
+\bigg(
        -\frac{1948}{81}
        +\frac{16 \zeta_2}{9}
\bigg) L^3
+\frac{44 L^4}{27}
\bigg\}
+ C_F^2 n_l T_F  \bigg\{
-\frac{2011}{81}
\nonumber\\ &
-\frac{64 c_1}{9}
-\frac{962 \zeta_2}{3}
+\frac{12232 \zeta _2^2}{45}
+\frac{2752 \zeta_3}{9}
-48 \zeta_3 \zeta_2
+40 \zeta _5
+\frac{448 \zeta_2 \ln (2)}{3}
+ \bigg(
        \frac{18812}{81}
        +\frac{682}{9} \zeta_2
\nonumber\\ &
        -\frac{392}{15} \zeta_2^2
        -\frac{1976}{9} \zeta_3
\bigg) L
+\bigg(
        -\frac{18817}{81}
        -\frac{100 \zeta_2}{3}
        +\frac{232 \zeta_3}{9}
\bigg) L^2
+\bigg(
        \frac{2032}{27}
        +\frac{58 \zeta_2}{9}
\bigg) L^3
-\frac{355 L^4}{27}
\nonumber\\ &
+L^5
\bigg\}
+ C_F n_l^2 T_F^2 \bigg\{
-\frac{29344}{729}
-\frac{976 \zeta_2}{81}
+\frac{928 \zeta_2^2}{45}
+\frac{256 \zeta_3}{9}
+\bigg(
        \frac{39352}{729}
        +\frac{608 \zeta_2}{27}
        +\frac{64 \zeta_3}{27}
\bigg) L
\nonumber\\ &
- \bigg(
        \frac{1624}{81}
        +\frac{32 \zeta_2}{9}
\bigg) L^2
+\frac{304 L^3}{81}
-\frac{8 L^4}{27}
\bigg\}
% 
%%%%%%%%%%%%%%%%%%%%%%%%%
\bigg] \,.
%%%%%%%%%%%%%%%%%%%%%%%%%%
% \end{align}
%%%%%%%%%%%%%%%%%%%%%%%%%%%%%%%%%%%%%%%%%%% Scalar
% 
%\begin{align}
\\ 
F_{S}^{(3)} &=
\frac{1}{\ep^3} \bigg[
%%%%%%%
-C_F^3 \frac{4}{3} (1-L)^3
%%%%%%%
- C_F^2 C_A \frac{22}{3}(1-L)^2
- C_F C_A^2 \frac{242}{27}(1-L)
+ C_F C_A n_l T_F \frac{176}{27}(1-L)
\nonumber\\&
+ C_F^2 n_l T_F \frac{8}{3} (1-L)^2
- C_F n_l^2 T_F^2 \frac{32}{27}(1-L)
\bigg]
%%%%%%%%%%%%%%%%%%%%%%%%%%
\nonumber\\&
+ \frac{1}{\ep^2} \bigg[
%%%%%%%
C_F^3 \bigg\{
-4
+4 \zeta_2
+\Big(
        8
        -8 \zeta_2
\Big) L
+\Big(
        -6
        +4 \zeta_2
\Big) L^2
+4 L^3
-2 L^4 
\bigg\}
%%%%%%%
+ C_F^2 C_A \bigg\{
\frac{32}{9}
+\frac{10}{3} \zeta_2
\nonumber\\ &
+4 \zeta_3
+\bigg(
        -\frac{166}{9}
        +\frac{2 \zeta_2}{3}
        -4 \zeta_3
\bigg) L
+\bigg(
        \frac{101}{9}
        -4 \zeta_2
\bigg) L^2
+\frac{11 L^3}{3}  
\bigg\}
+ C_F C_A^2 \bigg\{
\frac{1690}{81}
+\frac{44}{9} (\zeta_3-\zeta_2)
\nonumber\\ &
+\bigg(
        -\frac{2086}{81}
        +\frac{44 \zeta_2}{9}
\bigg) L
\bigg\}
+ C_F C_A n_l T_F \bigg\{
-\frac{1192}{81}
+\frac{16}{9} \zeta_2
-\frac{16}{9} \zeta_3
+\bigg(
        \frac{1336}{81}
        -\frac{16 \zeta_2}{9}
\bigg) L
\bigg\}
\nonumber\\ &
+ C_F^2 n_l T_F \bigg\{
-\frac{40}{9}
-\frac{8}{3} \zeta_2
+\bigg(
        \frac{80}{9}
        +\frac{8 \zeta_2}{3}
\bigg) L
-\frac{28 L^2}{9}
-\frac{4 L^3}{3} 
\bigg\}
+ C_F n_l^2 T_F^2 
\frac{160}{81}(1-L)
\bigg]
%%%%%%%%%%%%%%%%%%%%%%%%%%
\nonumber\\ &
+ \frac{1}{\ep} \bigg[
%%%%%%%
C_F^3 \bigg\{
-50
-10 \zeta_2
+\frac{236}{5} \zeta_2^2
+104 \zeta_3
+\bigg(
        62
        +34 \zeta_2
        -\frac{236}{5} \zeta_2^2
        -160 \zeta_3
\bigg) L
-\Big(
        20
        +18 \zeta_2
\nonumber\\ &
        -56 \zeta_3
\Big) L^2
+\bigg(
        \frac{26}{3}
        -6 \zeta_2
\bigg) L^3
-\frac{7 L^4}{3}
+\frac{5 L^5}{3} 
\bigg\}
%%%%%%%
+ C_F^2 C_A \bigg\{
\frac{1636}{27}
-59 \zeta_2
+\frac{26}{5} \zeta_2^2
-\frac{140}{3} \zeta_3
-4 \zeta_2 \zeta_3
\nonumber\\ &
+\bigg(
        -\frac{2030}{27}        
        +\frac{134}{3} \zeta_2
        -\frac{26}{5} \zeta_2^2
        +\frac{308}{3} \zeta_3
\bigg) L
+\bigg(
        \frac{835}{27}
        +\frac{37 \zeta_2}{3}
        -50 \zeta_3
\bigg) L^2
+\bigg(
        -\frac{212}{9}
        +6 \zeta_2
\bigg) L^3
\nonumber\\ &
+\frac{11 L^4}{9}  
\bigg\}
+ C_F C_A^2 \bigg\{
-\frac{343}{27}
+\frac{608}{27} \zeta_2
-\frac{88}{15} \zeta_2^2
-\frac{740}{27} \zeta_3
-\frac{8}{3} \zeta_2 \zeta_3
+12 \zeta_5
+\bigg(
        \frac{245}{9}
        -\frac{536}{27} \zeta_2
\nonumber\\ &
        +\frac{88}{15} \zeta_2^2
        +\frac{44}{9} \zeta_3
\bigg) L
\bigg\}
+ C_F C_A n_l T_F \bigg\{
\frac{356}{81}
-\frac{160}{27} \zeta_2
+\frac{496}{27} \zeta_3
+\bigg(
        -\frac{836}{81}
        +\frac{160 \zeta_2}{27}
        -\frac{112 \zeta_3}{9}
\bigg) L 
\bigg\}
\nonumber\\ &
% %%%%%%%%%%
+ C_F^2 n_l T_F \bigg\{
-\frac{38}{27}
+4 \zeta_2
-\frac{16}{3} \zeta_3
+\bigg(
        \frac{190}{27}
        +\frac{8 \zeta_2}{3}
        +\frac{16 \zeta_3}{3}
\bigg) L
+\bigg(
        -\frac{332}{27}
        -\frac{20 \zeta_2}{3}
\bigg) L^2
+\frac{64 L^3}{9}
\nonumber\\&
-\frac{4 L^4}{9}
\bigg\}
%%%%%%
+ C_F n_l^2 T_F^2 
\frac{32}{81}(1-L)
\bigg]
%%%%%%%%%%%%%%%%%%%%%%%%%%
+  \bigg[
X_{S,3}^{0,0}
+N_C^3 \bigg\{
-\frac{133949}{2916}
+\frac{6623 \zeta_2}{108}
-\frac{545 \zeta_2^2}{12}
+\frac{8156 \zeta_2^3}{315}
\nonumber\\ &
+\frac{3403 \zeta_3}{324}
-\frac{16 \zeta_3^2}{3}
+\frac{397 \zeta_3 \zeta_2}{6}
-\frac{830 \zeta_5}{3}
\bigg\}
%%%%%%%%%%
+C_F^3  \bigg\{
\bigg(
        -131
        +14 \zeta_2
        +528 \ln (2) \zeta_2
        -\frac{616}{5} \zeta_2^2
\nonumber\\ &
        -428 \zeta_3
        -8 \zeta_2 \zeta_3
        -276 \zeta_5
\bigg) L
- \bigg(
         87
        +79 \zeta_2
        -\frac{302}{5} \zeta_2^2
        -244 \zeta_3
\bigg) L^2
+\bigg(
        \frac{76}{3}
        +\frac{98 \zeta_2}{3}
        -\frac{268 \zeta_3}{3}
\bigg) L^3
\nonumber\\ &
+\bigg(
        -\frac{17}{2}
        +\frac{16 \zeta_2}{3}
\bigg) L^4
+L^5
-L^6
 \bigg\}
% 
%%%%%%%%%%
% 
+ C_F^2 C_A  \bigg\{
\bigg(
        -\frac{8569}{81}
        -\frac{1091}{9} \zeta_2
        -264 \ln (2) \zeta_2
        +\frac{2524}{15} \zeta_2^2
\nonumber\\ &
        +\frac{2686}{3} \zeta_3
        -46 \zeta_2 \zeta_3
        -194 \zeta_5
\bigg) L
+\bigg(
        \frac{9824}{81}
        +\frac{175}{6} \zeta_2
        -\frac{143}{5} \zeta_2^2
        -\frac{2312}{9} \zeta_3
\bigg) L^2
+\bigg(
        -\frac{1208}{27}
\nonumber\\ &
        -\frac{259 \zeta_2}{18}
        +\frac{232 \zeta_3}{3}
\bigg) L^3
+\bigg(
        \frac{2309}{108}
        -\frac{16 \zeta_2}{3}
\bigg) L^4
-\frac{11 L^5}{4}
 \bigg\}
% 
%%%%%%%%%%
% 
+ C_F C_A^2  \bigg\{
\bigg(
        \frac{10289}{1458}
        +\frac{9644}{81} \zeta_2
        -\frac{94}{3} \zeta_2^2
\nonumber\\ &
        -\frac{13900}{27} \zeta_3
        +\frac{88}{3} \zeta_2 \zeta_3
        +136 \zeta_5
\bigg) L
+\bigg(
        -\frac{11939}{162}
        +\frac{26}{9} \zeta_2        
        -\frac{44}{5} \zeta_2^2
        +88 \zeta_3
\bigg) L^2
+\bigg(
        \frac{1780}{81}
\nonumber\\ &
        -\frac{44 \zeta_2}{9}
\bigg) L^3
-\frac{121 L^4}{54}
\bigg\}
% 
%%%%%%%%%%
% 
+ C_F C_A n_l T_F \bigg\{
 \frac{35626}{729}
-\frac{2680 \zeta_2}{81}
-\frac{1340 \zeta_2^2}{9}
+160 \zeta_2 \ln (2)
-\frac{19348 \zeta_3}{81}
\nonumber\\ &
+\frac{8 \zeta_3 \zeta_2}{3}
+\frac{596 \zeta_5}{3}
+\bigg(
        -\frac{14998}{729}
        -\frac{6544}{81} \zeta_2
        +\frac{88}{15} \zeta_2^2
        +\frac{1448}{9} \zeta_3
\bigg) L
+\bigg(
        \frac{3454}{81}
        +\frac{32 \zeta_2}{3}
        -16 \zeta_3
\bigg) L^2
\nonumber\\ &
+\bigg(
        -\frac{1156}{81}
        +\frac{16 \zeta_2}{9}
\bigg) L^3
+\frac{44 L^4}{27}
\bigg\}
% 
%%%%%%%%%%
% 
+ C_F^2 n_l T_F  \bigg\{
 \frac{7727}{81}
+22 \zeta_2
+\frac{1592}{9} \zeta_2^2
-320 \zeta_2 \ln (2)
\nonumber\\ &
+\frac{3328}{9} \zeta_3
-48 \zeta_3 \zeta_2
+40 \zeta_5
+\bigg(
        \frac{1415}{81}
        +\frac{820}{9} \zeta_2
        -\frac{392}{15} \zeta_2^2
        -\frac{1904}{9} \zeta_3
\bigg) L
+\bigg(
        -\frac{2635}{81}
        -\frac{58 \zeta_2}{3}
\nonumber\\ &
        +\frac{232 \zeta_3}{9}
\bigg) L^2
+\bigg(
        \frac{460}{27}        
        +\frac{58 \zeta_2}{9}
\bigg) L^3
-\frac{175 L^4}{27}
+L^5
 \bigg\}
% 
%%%%%%%%%%
% 
+ C_F n_l^2 T_F^2  \bigg\{
-\frac{112}{729}
-\frac{2272 \zeta _2}{81}
+\frac{928 \zeta _2^2}{45}
\nonumber\\ &
+\frac{64 \zeta _3}{9}
+\bigg(
        \frac{3712}{729}
        +\frac{320 \zeta_2}{27}
        +\frac{64 \zeta_3}{27}
\bigg) L
+\bigg(
        -\frac{400}{81}
        -\frac{32 \zeta_2}{9}
\bigg) L^2
+\frac{160 L^3}{81}
-\frac{8 L^4}{27}
\bigg\}
\bigg] \,.
%%%%%%%%%%%%%%%%%%%%%%%%%%
\end{align}

%--------------------------------------------------------------------------------------------------------
Here we used the abbreviation
%--------------------------------------------------------------------------------------------------------
\begin{eqnarray}
%%B_4 &=& - 4 \zeta_2 \ln^2(2) + \frac{2}{3} \ln^4(2) - \frac{13}{2} \zeta_4 + 16 
%%\Li_4\left(\frac{1}{2}\right)
%%\\
c_1 &=& 12 \zeta_2 \ln^2(2) + \ln^4(2) + 24 \Li_4\left(\frac{1}{2}\right),
\end{eqnarray}
%--------------------------------------------------------------------------------------------------------
where $\Li_k(x)$ denotes the polylogarithm \cite{DUDE,LEWIN1,LEWIN2}. The functions $X_{V,3}^{0,0}$ and 
$X_{S,3}^{0,0}$ are independent of $L$ and contain the sub--leading contributions in $N_C$ which can be 
obtained through an exact calculation only, including the non-planar color topologies. The leading pole 
contributions are the same in the vector and scalar cases, also for the four--loop terms given below.

At the four--loop level, we obtain the complete contributions to the $\frac{1}{\ep^4}, \frac{1}{\ep^3}$ and 
$\frac{1}{\ep^2}$ poles, leading color contributions to $L$ and complete $L^2 \ldots L^{\rm Max}$ of the single 
pole in $\ep$ and leading color contributions to $L^2$ and complete $L^3 \ldots L^{\rm Max}$ for the finite 
pieces. The yet unknown terms are denoted by $X_{I,n}^{k,j}$ indicating the $\ep^k L^j$--coefficient of 
$F_{I}^{(n)}$. Again the terms $X_{I,n}^{k,j}$ do not contain logarithmic contributions.

A partial prediction of the four--loop vector and scalar form factor is given by
%--------------------------------------------------------------------------------------------------------
\begin{align}
F_V^{(4)} &=
\frac{1}{\varepsilon^4} \bigg[
         C_{F}^4  \frac{2}{3} (1-L)^4 
       + C_{A} C_{F}^3 \frac{22}{3} (1-L)^3
       + C_{A}^2 C_{F}^2 \frac{1331}{54} (1-L)^2
       + C_{A}^3 C_{F} \frac{1331}{54} (1-L)
\nonumber\\ &
       - C_{F}^3 n_l T_F \frac{8}{3} (1-L)^3
       - C_{A} C_{F}^2 n_l T_F \frac{484}{27} (1-L)^2
       - C_{A}^2 C_{F} n_l T_F \frac{242}{9} (1-L)
\nonumber\\ &
       + C_{F}^2 n_l^2 T_F^2  \frac{88}{27} (1-L)^2
       + C_{A} C_{F} n_l^2 T_F^2 \frac{88}{9} (1-L)
       - C_{F} n_l^3 T_F^3 \frac{32}{27}(1-L)
\bigg]
% 
%%%%%%%%%%%%%%%%%%%%%% % % %
% 
\nonumber\\ &
+ \frac{1}{\varepsilon^3}   \bigg[
         C_{F}^4   \bigg\{ \bigg( \frac{16}{3} - \frac{8}{3} \zeta_2 \bigg)
       + \bigg(  - 20 + 8 \zeta_2 \bigg) L
       + \bigg( \frac{88}{3} - 8 \zeta_2 \bigg) L^2
       + \bigg(  - \frac{64}{3} + \frac{8}{3} \zeta_2 \bigg) L^3
       + 8 L^4 
\nonumber\\ &
       - \frac{4}{3} L^5  \bigg\}
       + C_{A} C_{F}^3   \bigg\{ \bigg( \frac{166}{9} - 4 \zeta_3 - \frac{32}{3} \zeta_2 \bigg)
       + \bigg(  - 44 + 8 \zeta_3 + \frac{52}{3} \zeta_2 \bigg) L
       + \bigg( 40 - 4 \zeta_3 - \frac{8}{3} \zeta_2 \bigg) L^2
\nonumber\\ &
       - \bigg(\frac{196}{9} + 4 \zeta_2 \bigg) L^3
       + \frac{22}{3} L^4\bigg\}
       + C_{A}^2 C_{F}^2   \bigg\{ \bigg(  - \frac{2093}{81} - \frac{154}{9} \zeta_3 - \frac{22}{27} \zeta_2 \bigg)
       + \bigg( \frac{6298}{81} + \frac{154}{9} \zeta_3 
\nonumber\\ &
- \frac{440}{27} \zeta_2 \bigg) L
       + \bigg(  - \frac{3479}{81} + \frac{154}{9} \zeta_2 \bigg) L^2
       - \frac{242}{27} L^3 \bigg\}
       + C_{A}^3 C_{F}   \bigg\{ \bigg(  - \frac{12661}{162} - \frac{121}{9} \zeta_3 + \frac{121}{9} \zeta_2 \bigg)
\nonumber\\ &
       + \bigg( \frac{14839}{162} - \frac{121}{9} \zeta_2 \bigg) L \bigg\}
       + C_{F}^3 n_l T_F   \bigg\{ \bigg(  - \frac{8}{9} + \frac{16}{3} \zeta_2 \bigg)
       + \bigg( \frac{16}{3} - \frac{32}{3} \zeta_2 \bigg) L
       - \bigg( \frac{32}{3} - \frac{16}{3} \zeta_2 \bigg) L^2
\nonumber\\ &
       + \frac{80}{9} L^3 
       - \frac{8}{3} L^4 \bigg\}
       + C_{A} C_{F}^2  n_l T_F  \bigg\{ \bigg( \frac{2708}{81} + \frac{56}{9} \zeta_3 + \frac{184}{27} \zeta_2 \bigg)
       + \bigg(  - \frac{5260}{81} - \frac{56}{9} \zeta_3 
\nonumber\\ &
- \frac{16}{27} \zeta_2 \bigg) L
       + \bigg( \frac{2024}{81} - \frac{56}{9} \zeta_2 \bigg) L^2
       + \frac{176}{27} L^3 \bigg\}
       + C_{A}^2 C_{F} n_l T_F    \bigg\{ \bigg( \frac{730}{9} + \frac{88}{9} \zeta_3 - \frac{88}{9} \zeta_2 \bigg)
\nonumber\\ &
       +  \bigg(  - \frac{818}{9} + \frac{88}{9} \zeta_2 \bigg) L \bigg\}
       + C_{F}^2  n_l^2 T_F^2  \bigg\{ \bigg(  - \frac{608}{81} - \frac{64}{27} \zeta_2 \bigg)
       + \bigg( \frac{880}{81} + \frac{64}{27} \zeta_2 \bigg) L 
       - \frac{176}{81} L^2
\nonumber\\ &
       - \frac{32}{27}  L^3 \bigg\}
       + C_{A} C_{F} n_l^2 T_F^2   \bigg\{ \bigg(  - \frac{664}{27} - \frac{16}{9} \zeta_3 + \frac{16}{9} \zeta_2 \bigg)
       + \bigg( \frac{712}{27} - \frac{16}{9} \zeta_2 \bigg) L \bigg\}
\nonumber\\&
       + C_{F} n_l^3 T_F^3  \frac{160}{81} (1-L) 
\bigg]
% 
%%%%%%%%%%%%%%%%%%
% % 
+ \frac{1}{\varepsilon^2}   \bigg[
  C_{F}^4 \bigg\{
         \frac{212}{3} - \frac{232}{3} \zeta_3 + \frac{250}{3} \zeta_2 - \frac{236}{5} \zeta_2^2 - 96 \ln (2) \zeta_2
\nonumber\\&         
       + \bigg(  - \frac{551}{3} + 208 \zeta_3 - \frac{524}{3} \zeta_2 + \frac{472}{5} \zeta_2^2 + 192 \ln (2) \zeta_2 \bigg) L
       + \bigg( 185 - 184 \zeta_3 + 94 \zeta_2 - \frac{236}{5} \zeta_2^2 
\nonumber\\&       
- 96 \ln (2) \zeta_2 \bigg) L^2
       + \bigg(  - \frac{991}{9} + \frac{160}{3} \zeta_3 + \frac{8}{3} \zeta_2 \bigg) L^3
       + \bigg( 48 - \frac{16}{3} \zeta_2 \bigg) L^4
       - \frac{34}{3} L^5
       + \frac{13}{9} L^6
       \bigg\}
\nonumber\\ &
+ C_{A} C_{F}^3 \bigg\{
         \frac{188}{27} - 88 \zeta_3 
         + \frac{1613}{9} \zeta_2 + 8 \zeta_2 \zeta_3 - \frac{1436}{15} \zeta_2^2 - 
         128 \ln (2) \zeta_2
       + \bigg( \frac{851}{6} + 76 \zeta_3 - \frac{1987}{9} \zeta_2 
\nonumber\\&  
- 8 \zeta_2 \zeta_3 + \frac{1574}{15} 
         \zeta_2^2 + 80 \ln (2) \zeta_2 \bigg) L
       + \bigg(  - \frac{671}{3} + 60 \zeta_3 + \frac{290}{9} \zeta_2 - \frac{46}{5} \zeta_2^2 + 48 \ln (2) 
         \zeta_2 \bigg) L^2
\nonumber\\ &
       + \bigg( \frac{5057}{54} - 48 \zeta_3 + \frac{4}{3} \zeta_2 \bigg) L^3
       + \bigg(  - \frac{305}{18} + 8 \zeta_2 \bigg) L^4
       - \frac{11}{6} L^5
       \bigg\}
+ C_{A}^2 C_{F}^2 \bigg\{
       - \frac{10159}{54} + \frac{4700}{27} \zeta_3 
\nonumber\\ &       
       + 2 \zeta_3^2        
- \frac{1249}{81} \zeta_2 + \frac{100}{9} \zeta_2 
         \zeta_3 + \frac{79}{45} \zeta_2^2 + 88 \ln (2) \zeta_2 - 24 \zeta_5
       + \bigg( \frac{47179}{162} - \frac{7778}{27} \zeta_3 
       + \frac{6595}{81} \zeta_2 
\nonumber\\ &       
       - \frac{4}{3} \zeta_2 \zeta_3 
       - \frac{697}{45} \zeta_2^2 - 88 \ln (2) \zeta_2 + 24 \zeta_5 \bigg) L
       + \bigg(  - \frac{8159}{54} + \frac{902}{9} \zeta_3 - \frac{572}{9} \zeta_2 + \frac{206}{15} \zeta_2^2 \bigg) L^2
\nonumber\\ &       
       + \bigg( \frac{9925}{162} 
       - \frac{110}{9} \zeta_2 \bigg) L^3
          - \frac{121}{81} L^4  
       \bigg\}
+ C_{A}^3 C_{F}  \bigg\{
         \frac{4981}{54} + \frac{2341}{27} \zeta_3 - \frac{1978}{27} \zeta_2 + \frac{22}{3} \zeta_2 \zeta_3 + \frac{242}{15} 
         \zeta_2^2 
\nonumber\\ &         
         - 33 \zeta_5
       + \bigg(  - \frac{287}{2} - \frac{121}{9} \zeta_3 
       + \frac{1780}{27} \zeta_2 - \frac{242}{15} \zeta_2^2 \bigg) L
       \bigg\}
+ C_{F}^3 n_l T_F \bigg\{
       - \frac{673}{27} + \frac{208}{3} \zeta_3 - \frac{652}{9} \zeta_2 
\nonumber\\ & 
       + \frac{472}{15} \zeta_2^2 + 64 \ln (2) \zeta_2
       + \bigg( \frac{41}{9} 
- \frac{368}{3} \zeta_3 + \frac{740}{9} \zeta_2 
       - \frac{472}{15} \zeta_2^2 - 64 \ln (2) 
         \zeta_2 \bigg) L
       + \bigg( \frac{392}{9} + \frac{160}{3} \zeta_3 
\nonumber\\ &       
       + \frac{8}{9} \zeta_2 \bigg) L^2
       + \bigg(  - \frac{758}{27} - \frac{32}{3} \zeta_2 \bigg) L^3
       + \frac{38}{9} L^4
       + \frac{2}{3} L^5  
      \bigg\}
+ C_{A} C_{F}^2 n_l T_F \bigg\{
         \frac{5369}{81} - \frac{556}{9} \zeta_3 - \frac{1100}{81} \zeta_2 
\nonumber\\ &         
         - \frac{32}{9} \zeta_2 \zeta_3 + \frac{196}{45} 
         \zeta_2^2 - 32 \ln (2) \zeta_2       
       + \bigg(  - \frac{3883}{27} + \frac{1072}{9} \zeta_3 - \frac{1612}{81} \zeta_2 
       - \frac{196}{45} \zeta_2^2 + 32 \ln (2) \zeta_2 \bigg) L
\nonumber\\ &       
       + \bigg( \frac{9512}{81} - \frac{520}{9} \zeta_3 + \frac{880}{27} \zeta_2 \bigg) L^2
       + \bigg(  - \frac{3284}{81} + \frac{40}{9} \zeta_2 \bigg) L^3
       + \frac{88}{81}  L^4  
       \bigg\}
\nonumber\\ &
+ C_{A}^2 C_{F} n_l T_F \bigg\{
       - \frac{13241}{162} - \frac{2284}{27} \zeta_3 + \frac{1228}{27} \zeta_2 - \frac{8}{3} \zeta_2 \zeta_3 - \frac{88}{15} 
         \zeta_2^2 
+ 12 \zeta_5
       + \bigg( \frac{19313}{162} + \frac{352}{9} \zeta_3 
\nonumber\\ &       
       - \frac{1156}{27} \zeta_2 + \frac{88}{15} \zeta_2^2 \bigg) L
       \bigg\}
+ C_{F}^2 n_l^2 T_F^2 \bigg\{
         \frac{116}{81} - \frac{224}{27} \zeta_3 + \frac{464}{81} \zeta_2
       + \bigg( \frac{1132}{81} 
+ \frac{224}{27} \zeta_3 - \frac{80}{81} \zeta_2 \bigg) L
\nonumber\\ &
       + \bigg(  - \frac{1720}{81} - \frac{128}{27} \zeta_2 \bigg) L^2
       + \frac{488}{81} L^3  
       - \frac{16}{81} L^4 
       \bigg\}
+ C_{A} C_{F} n_l^2 T_F^2 \bigg\{
         \frac{1105}{81} + \frac{496}{27} \zeta_3 
- \frac{160}{27} \zeta_2
\nonumber\\ &
       + \bigg(  - \frac{1585}{81} - \frac{112}{9} \zeta_3 + \frac{160}{27} \zeta_2 \bigg) L
       \bigg\}
+ C_{F} n_l^3 T_F^3 \frac{32}{81} (1-L)
\bigg]
% 
%%%%%%%%%%%%%%%%%%%%%%%%%%%%%%%%%
% \nonumber\\&
% 
+ \frac{1}{\varepsilon}   \bigg[
X_{V,4}^{-1,0} + X_{V,4}^{-1,1} L
\nonumber\\ &
+ N_C^4   \bigg(  - \frac{2073967}{11664} - \frac{441}{2} \zeta_5 + \frac{178303}{648} \zeta_3 
- \frac{1}{3} \zeta_3^2
          + \frac{1141}{54} \zeta_2 + \frac{118}{9} \zeta_2 \zeta_3 - \frac{12142}{135} \zeta_2^2 
\nonumber\\&
+ \frac{1013}{45} 
         \zeta_2^3 \bigg) L
+ N_C^3 n_l T_F
         \bigg(  \frac{495349}{2916} + 82 \zeta_5 - \frac{2767}{18} \zeta_3 + \frac{2531}{162} \zeta_2 - \frac{32}{9} \zeta_2 \zeta_3 
         + \frac{694}{45} \zeta_2^2
         \bigg) L
\nonumber\\ &
+ C_{F}^4   \bigg\{
         \bigg( \frac{1972}{3} + 16 c_1 - 516 \zeta_5 - \frac{3244}{3} \zeta_3 + \frac{2974}{3} \zeta_2 + 8 
         \zeta_2 \zeta_3 - 504 \zeta_2^2 - 432 \ln (2) \zeta_2 \bigg) L^2
\nonumber\\ &
       + \bigg(  - \frac{3025}{6} + \frac{4928}{9} \zeta_3 
       - \frac{638}{3} \zeta_2 + \frac{1232}{15} \zeta_2^2 + 96
          \ln (2) \zeta_2 \bigg) L^3
       + \bigg( \frac{1942}{9} - \frac{992}{9} \zeta_3 + \frac{22}{3} \zeta_2 \bigg) L^4
\nonumber\\ &
       + \bigg(  - 62 + 6 \zeta_2 \bigg) L^5
       + \frac{32}{3}  L^6 
       - \frac{10}{9} L^7   
       \bigg\}
+ C_{A} C_{F}^3  \bigg\{
        \bigg(  - \frac{80599}{36} - 8 c_1 - 74 \zeta_5 + \frac{7045}{3} \zeta_3 
\nonumber\\&
- \frac{15098}{27} \zeta_2
          - 132 \zeta_2 \zeta_3 + \frac{760}{3} \zeta_2^2 + 216 \ln (2) \zeta_2 \bigg) L^2
       + \bigg( \frac{101510}{81} 
       - \frac{5404}{9} \zeta_3 + \frac{526}{9} \zeta_2 - 18 \zeta_2^2 
\nonumber\\ &
- 48 \ln (2)  \zeta_2 \bigg) L^3
       + \bigg(  - \frac{18415}{54} + \frac{305}{3} \zeta_3 + \frac{58}{9} \zeta_2 \bigg) L^4
       + \bigg( \frac{161}{3} - 9 \zeta_2 \bigg) L^5
       - \frac{22}{9} L^6
       \bigg\}
\nonumber\\ &
+ C_{A}^2 C_{F}^2  \bigg\{
         \bigg( \frac{3887225}{2916} + 260 \zeta_5 - \frac{99733}{81} \zeta_3 + \frac{14699}{81} \zeta_2 + 
         \frac{328}{3} \zeta_2 \zeta_3 + \frac{1232}{45} \zeta_2^2 \bigg) L^2
\nonumber\\ &         
       + \bigg(  
- \frac{190591}{486} + \frac{2024}{27} \zeta_3 
       + \frac{4829}{54} \zeta_2 - \frac{412}{15} 
         \zeta_2^2 \bigg) L^3
       + \bigg( \frac{38237}{972} - \frac{154}{27} \zeta_2 \bigg) L^4
       - \frac{605}{324} L^5   
       \bigg\}
\nonumber\\ &                                
+ C_{F}^3 n_l T_F \bigg\{
         \bigg( \frac{9925}{27} - \frac{2560}{9} \zeta_3 + \frac{4120}{27} \zeta_2 + \frac{104}{15} \zeta_2^2 \bigg) L^2
       + \bigg(  - \frac{26932}{81}        
                    - \frac{64}{9} \zeta_3 - \frac{428}{9} \zeta_2 \bigg) L^3
\nonumber\\ &
       + \bigg( \frac{2902}{27} + \frac{100}{9} \zeta_2 \bigg) L^4
       - \frac{52}{3} L^5 
       + \frac{8}{9}  L^6   
       \bigg\}
+ C_{A} C_{F}^2 n_l T_F \bigg\{
         \bigg(  - \frac{519812}{729} 
         + \frac{17068}{81} \zeta_3 - \frac{1912}{9} \zeta_2 
\nonumber\\ &
- \frac{208}{45} 
         \zeta_2^2 \bigg) L^2
       + \bigg( \frac{55448}{243}        
                    + \frac{416}{27} \zeta_3 - \frac{400}{27} \zeta_2 \bigg) L^3
       + \bigg(  - \frac{6446}{243} + \frac{56}{27} \zeta_2 \bigg) L^4
       + \frac{110}{81}  L^5 
       \bigg\}
\nonumber\\ &
+ C_{F}^2 n_l^2 T_F^2 \bigg\{
         \bigg(  - \frac{47536}{729} - \frac{1184}{81} \zeta_3 - \frac{1720}{27} \zeta_2 + \frac{3424}{135} \zeta_2^2
          \bigg) L          
       + \bigg( \frac{52316}{729} + \frac{512}{81} \zeta_3 + \frac{3568}{81} \zeta_2 \bigg) L^2
\nonumber\\ &
       + \bigg(  - \frac{6536}{243}        
       - \frac{104}{27} \zeta_2 \bigg) L^3
       + \frac{980}{243} L^4   
       - \frac{20}{81} L^5  
       \bigg\}
+ C_{A} C_{F} n_l^2 T_F^2 \bigg\{
         \bigg( \frac{923}{162} + \frac{1120}{27} \zeta_3          
         - \frac{304}{81} \zeta_2 
\nonumber\\&
- \frac{112}{15} \zeta_2^2 \bigg) L
       \bigg\}
+ C_{F} n_l^3 T_F^3 \bigg\{
         \bigg(  - \frac{32}{81} + \frac{64}{27} \zeta_3 \bigg) L
       \bigg\}
\bigg]
% 
%%%%%%%%%%%%%%%%%%%%%%%         Finite term
% 
+ \bigg[
X_{V,4}^{0,0} + X_{V,4}^{0,1} L + X_{V,4}^{0,2} L^2
+ N_C^4        
         \bigg( 
\nonumber\\ &
         \frac{201483937}{279936} - \frac{149}{3} \zeta_5 - \frac{308201}{486} \zeta_3          
         + \frac{1193}{12} \zeta_3^2 
               + \frac{2972341}{7776} \zeta_2 + \frac{839}{108} \zeta_2 \zeta_3 
- \frac{3451}{80} \zeta_2^2 + \frac{761}{36} \zeta_2^3 \bigg) L^2
\nonumber\\ &
+ N_C^3 n_l T_F 
         \bigg(  - \frac{2276657}{34992} + \frac{314}{3} \zeta_5 + \frac{7945}{243} \zeta_3 
         - \frac{93347}{486} \zeta_2 
               - \frac{55}{27} \zeta_2 \zeta_3 - \frac{12979}{540} \zeta_2^2 \bigg) L^2
+ C_{F}^4 \bigg\{
         \bigg(  
\nonumber\\ &
- \frac{16252}{9}          
                   - 16 c_1 + 1060 \zeta_5 + \frac{27328}{9} \zeta_3 - \frac{16273}{9} 
         \zeta_2 - \frac{1640}{9} \zeta_2 \zeta_3 
         + \frac{9596}{15} \zeta_2^2 + 368 \ln (2) \zeta_2 \bigg) L^3
\nonumber\\ &
       + \bigg( \frac{20759}{24} - 822 \zeta_3 + \frac{571}{2} \zeta_2        
       - \frac{299}{3} \zeta_2^2 - 56 \ln (2) \zeta_2 \bigg) L^4
       + \bigg(  - \frac{47299}{180} + \frac{1144}{9} \zeta_3 
       - \frac{38}{3} \zeta_2 \bigg) L^5
\nonumber\\ &
       + \bigg( \frac{2518}{45} - \frac{44}{9} \zeta_2 \bigg) L^6
       - \frac{343}{45} L^7  
       + \frac{27}{40} L^8 
       \bigg\}
+ C_{A} C_{F}^3 \bigg\{
         \bigg( \frac{3658325}{486} + 8 c_1 - 370 \zeta_5 - \frac{188444}{27} \zeta_3 
\nonumber\\ &
+ \frac{60428}{27} 
         \zeta_2 
         + 400 \zeta_2 \zeta_3 - \frac{60979}{90} \zeta_2^2 - \frac{728}{3} \ln (2) \zeta_2 \bigg) L^3         
       + \bigg(  - \frac{1588517}{648} + \frac{10637}{9} \zeta_3 - \frac{4825}{27} \zeta_2 
\nonumber\\ &
+ \frac{1531}{30} 
         \zeta_2^2 + 28 \ln (2) \zeta_2 \bigg) L^4
       + \bigg( \frac{176119}{360} - 116 \zeta_3 - 13 \zeta_2 \bigg) L^5      
       + \bigg(  - \frac{33517}{540} + \frac{22}{3} \zeta_2 \bigg) L^6
\nonumber\\ &       
       + \frac{209}{60}  L^7  
       \bigg\}
+ C_{A}^2 C_{F}^2 \bigg\{
         \bigg(  - \frac{4578454}{729} - 540 \zeta_5 
         + \frac{38435}{9} \zeta_3 - \frac{299959}{324} \zeta_2
          - \frac{1976}{9} \zeta_2 \zeta_3           
          + \frac{6752}{45} \zeta_2^2 
\nonumber\\ &          
          + \frac{88}{3} \ln (2) \zeta_2 \bigg) L^3
       + \bigg( \frac{1220761}{972} - \frac{6413}{18} \zeta_3 - \frac{691}{9} \zeta_2 + \frac{1486}{45} \zeta_2^2 \bigg) L^4
       - \bigg( \frac{1363133}{9720} - \frac{1001}{54} \zeta_2 \bigg) L^5
\nonumber\\ &       
       + \frac{3751}{540}  L^6  
       \bigg\}
+ C_{A}^3 C_{F} \bigg\{
         \bigg( \frac{467107}{486} - \frac{968}{3} \zeta_3 - \frac{694}{27} \zeta_2 + \frac{484}{15} \zeta_2^2 \bigg) L^3
       + \bigg(  - \frac{33803}{324} 
       + \frac{121}{9} \zeta_2 \bigg) L^4
\nonumber\\ &       
       + \frac{1331}{270}  L^5 
       \bigg\}
+ C_{F}^3 n_l T_F \bigg\{
         \bigg(  - \frac{427058}{243} + \frac{30664}{27} \zeta_3 - \frac{19252}{27} \zeta_2 + \frac{290}{9} 
         \zeta_2^2 + \frac{64}{3} \ln (2) \zeta_2 \bigg) L^3
\nonumber\\ &         
       + \bigg( \frac{115585}{162} 
- \frac{896}{9} \zeta_3 
+ \frac{2660}{27} \zeta_2 \bigg) L^4
       + \bigg(  - \frac{14269}{90} - 8 \zeta_2 \bigg) L^5
       + \frac{2807}{135}  L^6   
       - \frac{19}{15}  L^7   
       \bigg\}
\nonumber\\ &
+ C_{A} C_{F}^2 n_l T_F \bigg\{
         \bigg( \frac{2759932}{729} 
- \frac{98272}{81} \zeta_3 
+ \frac{64054}{81} \zeta_2 - \frac{160}{3} \zeta_2^2         
          - \frac{32}{3} \ln (2) \zeta_2 \bigg) L^3
\nonumber\\ &          
       + \bigg(  - \frac{21805}{27} + 78 \zeta_3 - \frac{1354}{81} \zeta_2 \bigg) L^4
       + \bigg( \frac{116653}{1215} - \frac{182}{27} \zeta_2 \bigg) L^5
       - \frac{682}{135}  L^6 
       \bigg\}
\nonumber\\ &
+ C_{A}^2 C_{F} n_l T_F \bigg\{
         \bigg(  - \frac{25553}{27} + 176 \zeta_3 - \frac{712}{27} \zeta_2 - \frac{176}{15} \zeta_2^2 \bigg) L^3
       + \bigg( \frac{2920}{27} 
       - \frac{88}{9} \zeta_2 \bigg) L^4
\nonumber\\ &       
        - \frac{242}{45} L^5
       \bigg\}
+ C_{F}^2 n_l^2 T_F^2 \bigg\{
         \bigg( \frac{2471390}{2187} - \frac{60640}{243} \zeta_3 + \frac{54284}{81} \zeta_2 + \frac{3596}{135} 
         \zeta_2^2 \bigg) L^2
\nonumber\\ &
       + \bigg(  - \frac{370744}{729} 
       + \frac{1616}{81} \zeta_3 
- \frac{12172}{81} \zeta_2 \bigg) L^3
       + \bigg( \frac{28904}{243} + \frac{1064}{81} \zeta_2 \bigg) L^4
       - \frac{18826}{1215} L^5       
       + \frac{124}{135}  L^6  
       \bigg\}
\nonumber\\ &
+ C_{A} C_{F} n_l^2 T_F^2 \bigg\{
\bigg(\frac{8996449}{2916} - \frac{2464}{9} \zeta_5 
         - \frac{86416}{81} \zeta_3 + \frac{459772}{243} \zeta_2
          + \frac{608}{9} \zeta_2 \zeta_3 + \frac{1232}{15} \zeta_2^2 \bigg) L
\nonumber\\ &
       + \bigg(  - \frac{308686}{243} + \frac{2032}{9} \zeta_3 
- \frac{11440}{27} \zeta_2 + \frac{80}{3} \zeta_2^2
          \bigg) L^2
       + \bigg( \frac{23302}{81} 
       - \frac{64}{3} \zeta_3 + \frac{736}{27} \zeta_2 \bigg) L^3
\nonumber\\ &
       + \bigg(  - \frac{968}{27} + \frac{16}{9} \zeta_2 \bigg) L^4
       + \frac{88}{45}  L^5   
       \bigg\}
+ C_{F} n_l^3 T_F^3 \bigg\{
         \bigg(  - \frac{558992}{2187} - \frac{320}{81} \zeta_3 - \frac{12992}{81} \zeta_2 - \frac{832}{45} \zeta_2^2
          \bigg) L
\nonumber\\&          
       + \bigg( \frac{79280}{729} + \frac{1216}{27} \zeta_2 \bigg) L^2       
       + \bigg(  - \frac{6496}{243} - \frac{128}{27} \zeta_2 \bigg) L^3
       + \frac{304}{81}  L^4  
       - \frac{32}{135}  L^5   
       \bigg\}
\bigg] \,.
\end{align}

%--------------------------------------------------------------------------------------------------------
% 
% 
\begin{align}
F_S^{4} &=
\frac{1}{\varepsilon^4} \bigg[
         C_{F}^4 \frac{2}{3}(1-L)^4
       + C_{A} C_{F}^3 \frac{22}{3} (1-L)^3
       + C_{A}^2 C_{F}^2  \frac{1331}{54}(1-L)^2
       + C_{A}^3 C_{F}  \frac{1331}{54}(1-L)
\nonumber\\ &
       - C_{F}^3 n_l T_F \frac{8}{3} (1-L)^3 
       - C_{A} C_{F}^2 n_l T_F  \frac{484}{27}(1-L)^2
       - C_{A}^2 C_{F} n_l T_F \frac{242}{9}(1-L)
\nonumber\\ &
       + C_{F}^2 n_l^2 T_F^2 \frac{88}{27} (1-L)^2
       + C_{A} C_{F} n_l^2 T_F^2  \frac{88}{9} (1-L)
       - C_{F} n_l^3 T_F^3 \frac{32}{27}(1-L)
\bigg]
% 
%%%%%%%%%%%%%%%%%%%%%% % % %
% 
\nonumber\\ &
+ \frac{1}{\varepsilon^3}   \bigg[
  C_{F}^4 \bigg\{
         \frac{8}{3} - \frac{8}{3} \zeta_2 
       - 8(1-\zeta_2)  L  
       + \bigg( \frac{28}{3} - 8 \zeta_2 \bigg) L^2
       + \bigg(  - \frac{20}{3} + \frac{8}{3} \zeta_2 \bigg) L^3
       +  4 L^4  
       - \frac{4}{3} L^5
       \bigg\}
\nonumber\\&       
+ C_{F}^3 C_{A} \bigg\{
         \frac{34}{9} - 4 \zeta_3 - \frac{32}{3} \zeta_2 
       + \bigg( \frac{22}{3} + 8 \zeta_3 + \frac{52}{3} \zeta_2 \bigg) L
       - \bigg( \frac{56}{3} + 4 \zeta_3 + \frac{8}{3} \zeta_2 \bigg) L^2
       + \bigg( \frac{2}{9} - 4 \zeta_2 \bigg) L^3
\nonumber\\ &
       + \frac{22}{3}  L^4 
       \bigg\}
+ C_{F}^2 C_{A}^2 \bigg\{
       - \frac{3545}{81} - \frac{154}{9} \zeta_3 - \frac{22}{27} \zeta_2 
       + \bigg( \frac{9928}{81} + \frac{154}{9} \zeta_3 - \frac{440}{27} \zeta_2 \bigg)L
       + \bigg(  - \frac{5657}{81} 
\nonumber\\ &
+ \frac{154}{9} \zeta_2 \bigg) L^2
       - \frac{242}{27} L^3
       \bigg\}
+ C_{F} C_{A}^3 \bigg\{
       - \frac{12661}{162} - \frac{121}{9} \zeta_3 + \frac{121}{9} \zeta_2 
       + \bigg( \frac{14839}{162} - \frac{121}{9} \zeta_2 \bigg) L
       \bigg\}
\nonumber\\ &
+ C_{F}^3 n_l T_F \bigg\{
         \frac{40}{9} + \frac{16}{3} \zeta_2 
       + \bigg(  - \frac{40}{3} - \frac{32}{3} \zeta_2 \bigg) L
       + \bigg( \frac{32}{3} + \frac{16}{3} \zeta_2 \bigg) L^2
       + \frac{8}{9}  L^3
       - \frac{8}{3} L^4  
       \bigg\}
\nonumber\\ &
+ C_{F}^2 C_{A} n_l T_F \bigg\{
         \frac{3764}{81} + \frac{56}{9} \zeta_3 + \frac{184}{27} \zeta_2 
       + \bigg(  - \frac{7900}{81} - \frac{56}{9} \zeta_3 - \frac{16}{27} \zeta_2 \bigg) L
       + \bigg( \frac{3608}{81} - \frac{56}{9} \zeta_2 \bigg) L^2
\nonumber\\ &
       + \frac{176}{27} L^3
       \bigg\}
+ C_{F} C_{A}^2 n_l T_F \bigg\{
         \frac{730}{9} + \frac{88}{9} \zeta_3 - \frac{88}{9} \zeta_2 
       + \bigg(  - \frac{818}{9} + \frac{88}{9} \zeta_2 \bigg) L
       \bigg\}
+ C_{F}^2 n_l^2 T_F^2 \bigg\{
       - \frac{800}{81} 
\nonumber\\ &
- \frac{64}{27} \zeta_2 
       + \bigg( \frac{1360}{81} + \frac{64}{27} \zeta_2 \bigg) L
       - \frac{464}{81} L^2   
       - \frac{32}{27} L^3  
       \bigg\}
+ C_{F} C_{A} n_l^2 T_F^2 \bigg\{
       - \frac{664}{27} - \frac{16}{9} \zeta_3 + \frac{16}{9} \zeta_2 
\nonumber\\&       
       + \bigg( \frac{712}{27} - \frac{16}{9} \zeta_2 \bigg) L
       \bigg\}
+ C_{F} n_l^3 T_F^3 \bigg\{
         \frac{160}{81} 
       - \frac{160}{81} L
       \bigg\}
\bigg]
% 
%%%%%%%%%%%%%%%%%%
% 
+ \frac{1}{\varepsilon^2}   \bigg[
  C_{F}^4 \bigg\{
         \frac{142}{3} - \frac{304}{3} \zeta_3 + \frac{28}{3} \zeta_2 - \frac{236}{5} \zeta_2^2 
\nonumber\\ &
       + \bigg(  - \frac{308}{3} + 256 \zeta_3 
       - \frac{128}{3} \zeta_2 + \frac{472}{5} \zeta_2^2 \bigg) L
       + \bigg( 70 - 208 \zeta_3 + 52 \zeta_2 - \frac{236}{5} \zeta_2^2 \bigg) L^2
       + \bigg(  - \frac{196}{9} 
\nonumber\\ &
+ \frac{160}{3} \zeta_3 - \frac{40}{3} \zeta_2 \bigg) L^3
       + \bigg( 9 - \frac{16}{3} \zeta_2 \bigg) L^4
       - \frac{10}{3} L^5  
       + \frac{13}{9}  L^6   
       \bigg\}
+ C_{F}^3 C_{A} \bigg\{
         \frac{545}{27} - 148 \zeta_3 + \frac{830}{9} \zeta_2 
\nonumber\\ &
+ 8 \zeta_2 \zeta_3 
- \frac{1436}{15} \zeta_2^2 
       + \bigg( \frac{431}{9} + 148 \zeta_3 - \frac{1798}{9} \zeta_2 - 8 \zeta_2 \zeta_3 + \frac{1574}{15} 
         \zeta_2^2 \bigg) L
       + \bigg(  - \frac{808}{9} + 48 \zeta_3 
\nonumber\\ &
+ \frac{776}{9} \zeta_2 
- \frac{46}{5} \zeta_2^2 \bigg) L^2
       + \bigg( \frac{1390}{27} - 48 \zeta_3 + \frac{40}{3} \zeta_2 \bigg) L^3
       + \bigg(  - \frac{503}{18} + 8 \zeta_2 \bigg) L^4
       -  \frac{11}{6} L^5
       \bigg\}
\nonumber\\ &
+ C_{F}^2 C_{A}^2 \bigg\{
       - \frac{13553}{162} - 24 \zeta_5 + \frac{3776}{27} \zeta_3 + 2 \zeta_3^2 + \frac{6374}{81} \zeta_2 + 
         \frac{100}{9} \zeta_2 \zeta_3 + \frac{79}{45} \zeta_2^2 
       + \bigg( \frac{254}{9} + 24 \zeta_5 
\nonumber\\ &
- \frac{6986}{27} \zeta_3 + \frac{160}{81} \zeta_2 
       - \frac{4}{3} \zeta_2 
         \zeta_3 - \frac{697}{45} \zeta_2^2 \bigg) L
       + \bigg( \frac{343}{18} + \frac{902}{9} \zeta_3 - \frac{704}{9} \zeta_2 + \frac{206}{15} \zeta_2^2 \bigg) L^2
       + \bigg( \frac{4418}{81} 
\nonumber\\ &
- \frac{110}{9} \zeta_2 \bigg) L^3
       - \frac{121}{81} L^4
       \bigg\}
+ C_{F} C_{A}^3 \bigg\{
         \frac{4981}{54} - 33 \zeta_5 + \frac{2341}{27} \zeta_3 - \frac{1978}{27} \zeta_2 + \frac{22}{3} \zeta_2 \zeta_3
          + \frac{242}{15} \zeta_2^2 
\nonumber\\ &
       + \bigg(  - \frac{287}{2} - \frac{121}{9} \zeta_3 + \frac{1780}{27} \zeta_2 - \frac{242}{15} \zeta_2^2 \bigg) L
       \bigg\}
+ C_{F}^3 n_l T_F \bigg\{
       - \frac{901}{27} + \frac{256}{3} \zeta_3 - \frac{184}{9} \zeta_2 + \frac{472}{15} \zeta_2^2 
\nonumber\\ &
       + \bigg( \frac{379}{9} - \frac{416}{3} \zeta_3 + \frac{344}{9} \zeta_2 - \frac{472}{15} \zeta_2^2 \bigg) L
       + \bigg( \frac{28}{9} + \frac{160}{3} \zeta_3 - \frac{64}{9} \zeta_2 \bigg) L^2
       + \bigg(  - \frac{560}{27} - \frac{32}{3} \zeta_2 \bigg) L^3
\nonumber\\ &
       + \frac{74}{9} L^4   
       + \frac{2}{3} L^5 
       \bigg\}
+ C_{F}^2 C_{A} n_l T_F \bigg\{
       - \frac{35}{9} - \frac{148}{3} \zeta_3 - \frac{3476}{81} \zeta_2 - \frac{32}{9} \zeta_2 \zeta_3 + \frac{196}{45} 
         \zeta_2^2 
       + \bigg( \frac{2219}{81} 
\nonumber\\ &
+ \frac{976}{9} \zeta_3 + \frac{332}{81} \zeta_2 - \frac{196}{45} \zeta_2^2 \bigg) L
       + \bigg( \frac{788}{81} - \frac{520}{9} \zeta_3 + \frac{1024}{27} \zeta_2 \bigg) L^2
       + \bigg(  - \frac{2888}{81} + \frac{40}{9} \zeta_2 \bigg) L^3
\nonumber\\ &
       + \frac{88}{81} L^4
       \bigg\}
+ C_{F} C_{A}^2 n_l T_F \bigg\{
       - \frac{13241}{162} + 12 \zeta_5 - \frac{2284}{27} \zeta_3 + \frac{1228}{27} \zeta_2 
- \frac{8}{3} \zeta_2 
         \zeta_3 - \frac{88}{15} \zeta_2^2 
\nonumber\\ &
       + \bigg( \frac{19313}{162} + \frac{352}{9} \zeta_3 
- \frac{1156}{27} \zeta_2 + \frac{88}{15} \zeta_2^2 \bigg) L
       \bigg\}
+ C_{F}^2 n_l^2 T_F^2 \bigg\{
         \frac{964}{81} - \frac{224}{27} \zeta_3 + \frac{320}{81} \zeta_2 
       - \bigg(  \frac{844}{81} 
\nonumber\\ &
- \frac{224}{27} \zeta_3 - \frac{64}{81} \zeta_2 \bigg) L
       + \bigg(  - \frac{520}{81} - \frac{128}{27} \zeta_2 \bigg) L^2
       + \frac{416}{81}  L^3   
       - \frac{16}{81} L^4  
       \bigg\}
+ C_{F} C_{A} n_l^2 T_F^2 \bigg\{
         \frac{1105}{81} + \frac{496}{27} \zeta_3 
\nonumber\\ &
- \frac{160}{27} \zeta_2 
       + \bigg(  - \frac{1585}{81} - \frac{112}{9} \zeta_3 + \frac{160}{27} \zeta_2 \bigg) L
       \bigg\}
+ C_{F} n_l^3 T_F^3 \bigg\{
         \frac{32}{81} 
       - \frac{32}{81} L
       \bigg\}
\bigg]
% 
%%%%%%%%%%%%%%%%%%%%%%%%%%%%%%%%%
% 
+ \frac{1}{\varepsilon}   \bigg[
X_{S,4}^{-1,0} + X_{S,4}^{-1,1} L
\nonumber\\ &
+ N_C^4  
         \bigg( \frac{598439}{11664} - \frac{429}{2} \zeta_5 + \frac{42851}{324} \zeta_3 - \frac{1}{3} \zeta_3^2 + 
           \frac{1327}{108} \zeta_2 + \frac{487}{9} \zeta_2 \zeta_3 - \frac{40243}{540} \zeta_2^2 + \frac{1013}{45} 
           \zeta_2^3 \bigg) L
\nonumber\\ &
+ N_C^3 n_l T_F 
         \bigg(  - \frac{117913}{5832} + 82 \zeta_5 - \frac{2021}{18} \zeta_3 + \frac{6017}{162} \zeta_2 
                  - \frac{32}{9} \zeta_2 \zeta_3 - \frac{317}{45} \zeta_2^2 \bigg) L
\nonumber\\ &
+ C_{F}^4 \bigg\{
         \bigg(  - \frac{293}{3} - 516 \zeta_5 - \frac{2068}{3} \zeta_3 
         + \frac{424}{3} \zeta_2 + 8 \zeta_2 
         \zeta_3 - 168 \zeta_2^2 + 528 \ln(2) \zeta_2 \bigg) L^2
       + \bigg(  - 118 
\nonumber\\ &
+ \frac{3560}{9} \zeta_3 - \frac{388}{3} \zeta_2 + \frac{1232}{15} \zeta_2^2 \bigg) L^3
       + \bigg( \frac{271}{9} - \frac{992}{9} \zeta_3 + \frac{118}{3} \zeta_2 \bigg) L^4
       + \bigg(  - \frac{29}{3} + 6 \zeta_2 \bigg) L^5
       + 2 L^6
\nonumber\\ &   
       - \frac{10}{9} L^7   
       \bigg\}
- C_{F}^3 C_{A} \bigg\{
         \bigg(   \frac{5315}{27} + 74 \zeta_5 - \frac{5002}{3} \zeta_3 + \frac{8681}{27} \zeta_2 + 132 
         \zeta_2 \zeta_3 - \frac{2342}{15} \zeta_2^2 
         + 264 \ln (2) \zeta_2 \bigg) L^2
\nonumber\\ &
       - \bigg( \frac{16982}{81} - \frac{3244}{9} \zeta_3 - \frac{140}{9} \zeta_2 - 18 \zeta_2^2 \bigg) L^3
       + \bigg( \frac{4225}{54} - \frac{305}{3} \zeta_3 + \frac{158}{9} \zeta_2 \bigg) L^4
       - \bigg( \frac{223}{6} - 9 \zeta_2 \bigg) L^5
\nonumber\\ &
       + \frac{22}{9} L^6
       \bigg\}
+ C_{F}^2 C_{A}^2 \bigg\{
         \bigg( \frac{96527}{729} + 260 \zeta_5 - \frac{90526}{81} \zeta_3 + \frac{30721}{162} \zeta_2 + 
         \frac{328}{3} \zeta_2 \zeta_3 + \frac{88}{9} \zeta_2^2 \bigg) L^2
\nonumber\\ &
       + \bigg(  - \frac{53009}{243} + \frac{2024}{27} \zeta_3 + \frac{4631}{54} \zeta_2 - \frac{412}{15} \zeta_2^2
          \bigg) L^3
       + \bigg( \frac{27347}{972} - \frac{154}{27} \zeta_2 \bigg) L^4
       - \frac{605}{324} L^5
       \bigg\}
\nonumber\\ &
+ C_{F}^3 n_l T_F \bigg\{
         \bigg(  - \frac{1018}{27} - \frac{3280}{9} \zeta_3 + \frac{4084}{27} \zeta_2 + \frac{104}{15} \zeta_2^2 \bigg) L^2
       + \bigg(  - \frac{2272}{81} - \frac{64}{9} \zeta_3 - \frac{176}{9} \zeta_2 \bigg) L^3
\nonumber\\ &
       + \bigg( \frac{754}{27} + \frac{100}{9} \zeta_2 \bigg) L^4
       - \frac{34}{3} L^5  
       + \frac{8}{9}  L^6 
       \bigg\}
+ C_{F}^2 C_{A} n_l T_F \bigg\{
         \bigg(  - \frac{59948}{729} + \frac{24088}{81} \zeta_3 - \frac{524}{3} \zeta_2 
\nonumber\\ &    
- \frac{208}{45} \zeta_2^2
          \bigg) L^2
   + \bigg( \frac{27944}{243} + \frac{416}{27} \zeta_3 - \frac{364}{27} \zeta_2 \bigg) L^3
       + \bigg(  - \frac{4466}{243} + \frac{56}{27} \zeta_2 \bigg) L^4
       + \frac{110}{81}  L^5   
       \bigg\}
\nonumber\\ &
+ C_{F}^2 n_l^2 T_F^2 \bigg\{
         \bigg( \frac{18740}{729} - \frac{2624}{81} \zeta_3 - \frac{1840}{27} \zeta_2 + \frac{3424}{135} \zeta_2^2 \bigg) L
       + \bigg( \frac{5984}{729} + \frac{512}{81} \zeta_3 
       + \frac{2632}{81} \zeta_2 \bigg) L^2
\nonumber\\ &
       + \bigg(  - \frac{2576}{243} - \frac{104}{27} \zeta_2 \bigg) L^3
       + \frac{620}{243}  L^4  
       - \frac{20}{81}  L^5  
       \bigg\}
+ C_{F} C_{A} n_l^2 T_F^2 \bigg\{
         \bigg( \frac{923}{162} + \frac{1120}{27} \zeta_3 - \frac{304}{81} \zeta_2 
\nonumber\\ &
- \frac{112}{15} \zeta_2^2 \bigg) L 
       \bigg\}
+ C_{F} n_l^3 T_F^3 \bigg\{
         \bigg(  - \frac{32}{81} + \frac{64}{27} \zeta_3 \bigg) L 
       \bigg\}
\bigg]
% 
%%%%%%%%%%%%%%%%%%%%%%%                                                                       Finite term
% 
+ \bigg[
X_{S,4}^{0,0} + X_{S,4}^{0,1} L + X_{S,4}^{0,2} L^2
+ N_C^4 \bigg\{
         \bigg(  
\nonumber\\ &
- \frac{31524563}{69984} - \frac{455}{3} \zeta_5 + \frac{194209}{1944} \zeta_3 
         + \frac{1193}{12}
          \zeta_3^2 + \frac{26833}{486} \zeta_2 - \frac{187}{108} \zeta_2 \zeta_3 
          - \frac{7343}{120} \zeta_2^2
          + \frac{761}{36} \zeta_2^3 \bigg) L^2
       \bigg\}
\nonumber\\ &
+ N_C^3 n_l T_F \bigg\{
         \bigg( \frac{472810}{2187} + \frac{314}{3} \zeta_5 - \frac{23231}{243} \zeta_3 + \frac{23}{243} \zeta_2 
         - 
         \frac{55}{27} \zeta_2 \zeta_3 - \frac{271}{540} \zeta_2^2 \bigg) L^2
       \bigg\}
+ C_{F}^4 \bigg\{
         \bigg( \frac{239}{9} 
\nonumber\\ &
+ 1060 \zeta_5 + \frac{11132}{9} \zeta_3 - \frac{3964}{9} \zeta_2 - \frac{1640}{9} 
         \zeta_2 \zeta_3 + \frac{2852}{15} \zeta_2^2 - 528 \ln(2) \zeta_2 \bigg) L^3
       + \bigg( \frac{275}{2} - \frac{1388}{3} \zeta_3 
\nonumber\\ &
+ 171 \zeta_2        
       - \frac{299}{3} \zeta_2^2 \bigg) L^4
       + \bigg(  - \frac{1321}{45} + \frac{1144}{9} \zeta_3 - \frac{146}{3} \zeta_2 \bigg) L^5
       + \bigg( \frac{731}{90} - \frac{44}{9} \zeta_2 \bigg) L^6
       - \frac{43}{45} L^7  
\nonumber\\ & 
       + \frac{27}{40}  L^8  
       \bigg\}
+ C_{F}^3 C_{A} \bigg\{
         \bigg( \frac{133270}{243}          
         - 370 \zeta_5 - \frac{116066}{27} \zeta_3 + \frac{36173}{27} \zeta_2 + 
         400 \zeta_2 \zeta_3 - \frac{40567}{90} \zeta_2^2 
\nonumber\\ &
+ 264 \ln (2) \zeta_2 \bigg) L^3
       + \bigg(  - \frac{49319}{162} + \frac{6470}{9} \zeta_3 - \frac{3502}{27} \zeta_2        
       + \frac{1531}{30} \zeta_2^2
          \bigg) L^4
       + \bigg( \frac{7771}{90} - 116 \zeta_3 
\nonumber\\ &
+ 14 \zeta_2 \bigg) L^5
       + \bigg(  - \frac{16687}{540} + \frac{22}{3} \zeta_2 \bigg) L^6
       + \frac{209}{60}  L^7 
       \bigg\}
+ C_{F}^2 C_{A}^2 \bigg\{
         \bigg(  - \frac{411373}{729} - 540 \zeta_5          
         + \frac{28843}{9} \zeta_3 
\nonumber\\ &
- \frac{52132}{81} \zeta_2 - 
         \frac{1976}{9} \zeta_2 \zeta_3 + \frac{8336}{45} \zeta_2^2 \bigg) L^3
       + \bigg( \frac{376009}{972}   
       - \frac{6413}{18} \zeta_3 - \frac{383}{9} \zeta_2 + \frac{1486}{45} \zeta_2^2 \bigg) L^4
\nonumber\\ &
       + \bigg(  - \frac{203297}{2430}        
       + \frac{1001}{54} \zeta_2 \bigg) L^5
       + \frac{3751}{540}  L^6  
       \bigg\}
+ C_{F} C_{A}^3 \bigg\{
         \bigg( \frac{83618}{243} - \frac{968}{3} \zeta_3 
         - \frac{694}{27} \zeta_2 + \frac{484}{15} \zeta_2^2 \bigg) L^3
\nonumber\\ &
       + \bigg(  - \frac{5456}{81} + \frac{121}{9} \zeta_2 \bigg) L^4
       + \frac{1331}{270}  L^5  
       \bigg\}
+ C_{F}^3 n_l T_F \bigg\{
         \bigg(  - \frac{13454}{243} 
         + \frac{30448}{27} \zeta_3 - \frac{11980}{27} \zeta_2 
\nonumber\\ &
+ \frac{290}{9} 
         \zeta_2^2 \bigg) L^3
       + \bigg( \frac{5243}{81} - \frac{896}{9} \zeta_3 + \frac{1724}{27} \zeta_2 \bigg) L^4
       + \bigg(  - \frac{1352}{45} - 8 \zeta_2 \bigg) L^5
       + \frac{1277}{135}  L^6 
       - \frac{19}{15} L^7  
       \bigg\}
\nonumber\\ &
+ C_{F}^2 C_{A} n_l T_F \bigg\{
         \bigg( \frac{269308}{729} - \frac{87616}{81} \zeta_3 + \frac{40528}{81} \zeta_2 - \frac{160}{3} \zeta_2^2
          \bigg) L^3
       + \bigg(  - \frac{18701}{81} + 78 \zeta_3 
\nonumber\\ &
       - \frac{2362}{81} \zeta_2 \bigg) L^4
       + \bigg( \frac{66658}{1215} - \frac{182}{27} \zeta_2 \bigg) L^5
       - \frac{682}{135} L^6 
       \bigg\}
+ C_{F} C_{A}^2 n_l T_F \bigg\{
        \bigg(  - \frac{8693}{27} + 176 \zeta_3 - \frac{712}{27} \zeta_2 
\nonumber\\ &
- \frac{176}{15} \zeta_2^2 \bigg) L^3
       + \bigg( \frac{1831}{27} - \frac{88}{9} \zeta_2 \bigg) L^4
       - \frac{242}{45} L^5
       \bigg\}
+ C_{F}^2 n_l^2 T_F^2 \bigg\{
         \bigg(  - \frac{47206}{2187} - \frac{78928}{243} \zeta_3 + \frac{22760}{81} \zeta_2 
\nonumber\\ &
+ \frac{3596}{135} 
         \zeta_2^2 \bigg) L^2
       + \bigg(  - \frac{37744}{729} + \frac{1616}{81} \zeta_3 - \frac{7168}{81} \zeta_2 \bigg) L^3
       + \bigg( \frac{7256}{243} 
       + \frac{1064}{81} \zeta_2 \bigg) L^4
       - \frac{9736}{1215} L^5   
\nonumber\\ &
       + \frac{124}{135}  L^6 
       \bigg\}
+ C_{F} C_{A} n_l^2 T_F^2 \bigg\{
         \bigg(  - \frac{823055}{2916} + 96 \zeta_4 - \frac{2464}{9} \zeta_5 - \frac{68272}{81} \zeta_3 + 
         \frac{146572}{243} \zeta_2 
         + \frac{608}{9} \zeta_2 \zeta_3 
\nonumber\\ &
+ \frac{160}{3} \zeta_2^2 \bigg) L
       + \bigg(  - \frac{26293}{243} + \frac{2032}{9} \zeta_3 - \frac{6400}{27} \zeta_2 + \frac{80}{3} \zeta_2^2 \bigg) L^2
       + \bigg( \frac{7030}{81} - \frac{64}{3} \zeta_3 + \frac{736}{27} \zeta_2 \bigg) L^3
\nonumber\\ &
       + \bigg(  - \frac{572}{27} + \frac{16}{9} \zeta_2 \bigg) L^4
       + \frac{88}{45}  L^5  
       \bigg\}
+ C_{F} n_l^3 T_F^3 \bigg\{
         \bigg(  - \frac{20864}{2187} - \frac{320}{81} \zeta_3 - \frac{3200}{81} \zeta_2 - \frac{832}{45} \zeta_2^2 \bigg) L
\nonumber\\ &
       + \bigg( \frac{8000}{729} + \frac{640}{27} \zeta_2 \bigg) L^2
       + \bigg(  - \frac{1600}{243} - \frac{128}{27} \zeta_2 \bigg) L^3
       + \frac{160}{81}  L^4 
       - \frac{32}{135}  L^5   
       \bigg\}
\bigg] \,.
\end{align}
% 
% 

%--------------------------------------------------------------------------------------------------------
To obtain $X_{I,4}^{-1,0}$ for $I=V,S$ one needs the massive quark anomalous dimension at 
four--loop order, $\gamma_{Q}^{(3)}$, and the complete contributions from three--loop order 
are necessary as well. On the other hand, 
$X_{I,4}^{-1,1}$ lacks the 
non-planar contributions
only, which can be obtained once the massless cusp anomalous dimension $A_q^{(4)}$ is known completely. It similarly follows 
for
the remaining of the functions $X_{I,4}^{0,k}$.

The exact result for the color--planar and complete light quark contributions for the vector and scalar form 
factors are available \cite{Henn:2016tyf, Ablinger:2018yae, FORMF3, Lee:2018rgs} now. 
We successfully cross-checked our results in the corresponding limit.
An interesting point to note is that
similar to the coefficients $g_{I}^{n,k}$, the coefficients $\cC_{I}^{n,k}$ are also in accordance with principle 
of leading transcendentality, 
\textit{i.e.} the leading transcendental terms for each order in the $\ep$-expansion are the same in both the vector 
and scalar cases. This aspect is of importance considering form factors in super--symmetric $\cal N=$~4 SYM theories.

Finally, we would like to mention that the $Q^2$-dependent parts of $\hat{F}_I$ and the massless form factor 
$\overline{F}_I = F_I(m=0)$
are the same. The universal function $Z^{(m|0)}_I$ \cite{Mitov:2006xs} is then given by
 %--------------------------------------------------------------------------------------------------------
\begin{eqnarray}
Z^{(m|0)}_I\left(\frac{m^2}{\mu^2},a_s,\ep\right) &=& 
F_I\left(\frac{Q^2}{\mu^2},\frac{m^2}{\mu^2},a_s,\ep\right)
\left(\overline{F}_I\left(\frac{Q^2}{\mu^2},a_s,\ep\right)\right)^{-1} 
\nonumber\\ 
&=& {\cal C}_I(a_s,\ep) 
\hat{F}_I\left(\frac{Q^2}{\mu^2},\frac{m^2}{\mu^2},a_s,\ep\right)  
\left(\overline{F}_I\left(\frac{Q^2}{\mu^2},a_s,\ep\right)\right)^{-1}.
\end{eqnarray} 
%--------------------------------------------------------------------------------------------------------
%--------------------------------------------------------------------------------------------------------
\section{Conclusion}  
\label{sec:conclu}
%--------------------------------------------------------------------------------------------------------

\vspace*{1mm}
\noindent
We presented a systematic study of the massive form factors at $Q^2 \gg m^2$ at three--loop order by solving
the associated evolution equations both in the vector and scalar cases. The universal structure of the IR 
singularities, along with the interplay of the various anomalous dimensions, has enabled us to obtain all 
asymptotic  corrections at three--loop order for all logarithmic contributions. We also obtained partial 
four--loop results, still containing pieces, which can only be determined by performing a four--loop calculation. 
The Dirac axial-vector and vector form factor, and likewise the pseudo-scalar and scalar form factors, agree 
in this limit, while the Pauli vector form factors vanish. We remark that there are additional corrections due 
to massive internal quark loops, which have not been considered in the present paper and are the subject of 
further investigations. The present results constrain future calculations and may serve as important checks.
%--------------------------------------------------------------------------------------------------------

\vspace*{3mm}
\noindent
{\bf Acknowledgment.}\\
We thank would like to thank J. Ablinger, T. Ahmed, S. Moch, V. Ravindran and C. Schneider 
for discussions. This work has been funded in part by EU TMR network SAGEX agreement 
No. 764850 (Marie 
Sk\l{}odowska-Curie) and COST action CA16201: Unraveling new physics 
at the LHC through the precision frontier. 
%--------------------------------------------------------------------------------------------------------

\end{document}